\documentclass[fleqn,usenatbib]{mnras}
\usepackage{newtxtext,newtxmath}
\usepackage[T1]{fontenc}
\usepackage{ae,aecompl}

\usepackage{hyperref}
\usepackage{booktabs}
\usepackage{multirow}
\usepackage{pdflscape}
\usepackage{dblfloatfix}
\usepackage{float}
\usepackage[labelfont=bf]{caption}
\usepackage{graphicx}	
\usepackage{amsmath}	
\usepackage{amssymb}	

\newcommand{\kms}{\,km\,s$^{-1}$}

\title[Spectral classification of S0 galaxies in the nearby universe]{The local universe in the era of large surveys. I. Spectral classification of S0 galaxies}

\author[J. L. Tous et al.]{
J. L. Tous$^{1,2}$\thanks{E-mail: jtous@fqa.ub.edu (JLT)},
J. M. Solanes$^{1,2}$
and J. D. Perea$^{3}$
\\
$^{1}$Departament de F\'isica Qu\`antica i Astrof\'isica, Universitat de Barcelona, C. Mart\'i i Franqu\`es 1, 08028 Barcelona, Spain\\
$^{2}$Institut de Ci\`encies del Cosmos (ICCUB), Universitat de Barcelona., C. Mart\'i i Franqu\`es 1, 08028 Barcelona, Spain\\
$^{3}$Departamento de Astronom\'ia Extragal\'actica, Instituto de Astrof\'isica de Andaluc\'ia, IAA--CSIC, Glorieta de la Astronom\'ia s/n,18008 Granada, Spain
}

\date{18 May 2020}
\pubyear{2020}

\begin{document}
\label{firstpage}
\pagerange{\pageref{firstpage}--\pageref{lastpage}}
\maketitle

\begin{abstract}
This is the first paper in a series devoted to review the main properties of galaxies designated S0 in the Hubble classification system. Our aim is to gather abundant and, above all, robust information on the most relevant physical parameters of this poorly-understood morphological type and their possible dependence on the environment that could later be used to assess their possible formation channel(s). The adopted approach combines the characterisation of the fundamental features of the optical spectra of $68{,}043$ S0 with heliocentric $z\lesssim 0.1$ with the exploration of a comprehensive set of their global attributes. A principal component analysis is used to reduce the huge number of dimensions of the spectral data to a low-dimensional space facilitating a bias-free machine-learning-based classification of the galaxies. This procedure has revealed that objects bearing the S0 designation consist, despite their similar morphology, of two separate sub-populations with statistically inconsistent physical properties. Compared to the absorption-dominated S0, those with significant nebular emission are, on average, somewhat less massive, more luminous with less concentrated light profiles, have a younger, bluer and metal-poorer stellar component, and avoid high-galaxy-density regions. Noteworthy is the fact that the majority of members of this latter class, which accounts for at least a quarter of the local S0 population, show star formation rates and spectral characteristics entirely similar to those seen in late spirals. Our findings suggest that star-forming S0 might be less rare than hitherto believed and raise the interesting possibility of identifying them with plausible progenitors of their quiescent counterparts.
\end{abstract}




\section{Introduction}
\label{S:intro}

\citet{Hubble+1936} introduced in his book The Realm of Nebulae a hypothetical class of S0 galaxies that were supposed to have intermediate characteristics between those of E (elliptical) and S (spiral) galaxies. Later works by Hubble himself and other eminent morphologists \citetext{see, for example, the review on the early history of galaxy classification by \citealt{Sandage2005}} indicated that the lens-shaped  galaxies that were often found in galaxy aggregations had to be identified as the S0 postulated by Hubble. This identification seemed reasonable because (typical) lenticular galaxies (from now on the terms S0 and lenticular will be used indifferently) contain both a significant spheroidal central component and a disc component. Although these two structural elements are present to a greater or lesser extent in most spirals -- the exception being some bulgeless discs --, the main difference between them and lenticulars is that the objects of the latter class host thick and smooth discs, without spiral structure and that consist mainly of aging stars.

Fifteen years later, the observation that S0, unlike in the field, comprise the dominant population in the central regions of many rich clusters of galaxies led \citet{Spitzer&Baade1951} to suggest that such objects were, in fact, spirals from which the interstellar gas has been removed by collisions with other galaxies, quenching the star formation in the discs. It would take twenty-one more years to realise that it was much more likely that the gas-sweeping mechanism operating in these large structures is actually the ram pressure produced by the intracluster medium upon the discs as galaxies move through the cluster \citep{Gunn&Gott1972}.

The view that the cluster environment can drive the transformation of S into S0 is supported by several pieces of observational evidence. One of the most important is the morphology-density relation in rich clusters \citep{Dressler1980,PG84,GHC86,Got03,Cap11,Hou15}, which convincingly demonstrates that in such systems the number fraction of S0 rises with local projected density at the same pace that the S+Irr (irregular) fraction diminishes. Yet, the most direct evidence that late-type galaxies (LTG) may evolve in dense environments to become S0 is what is generically known as the Butcher-Oemler effect \citep{Butcher&Oemler+1978} consisting in the observation that a number of moderately high-redshift ($z\sim 0.4$ -- $0.5$) clusters contain a higher proportion of blue, star-forming objects, as well as a factor two lower S0 and E fractions than local galaxy associations  \citep[e.g.][]{Cou94,Dressler+1997,Couch+1998,Poggianti+1999,Fasano+2000,Tre03}.
  
In spite of being the dominant constituent in the inner regions of many rich clusters, S0 also occur, albeit less frequently -- in terms of the relative population fractions, but not necessarily in terms of global numbers --, in small groups and even in the general field, where one can hardly expect that hydrodynamic interactions between the interstellar medium (ISM) and the much more tenuous intergalactic gas account for their formation. The presence of  bulge-enhanced, spiral-less disc galaxies in lower-density environments is explained instead by invoking close gravitational interactions between pairs of late-type objects that may lead to their merger \citep{Barnes+1999, Querejeta+2015, Eliche-Moral+2018}, since this sort of transformation mechanism is most effective when the relative speeds of the galaxies are low and similar to the internal velocities of their stars.
 
We are therefore confronted with the possibility that galaxies that fit into Hubble's S0 class may have followed at least two distinct formation pathways: some could be descendants of evolved S and/or Irr whose gaseous component is swept by the dense intracluster medium and their shapes altered by tidal interactions either with the global cluster potential, or with other cluster members, or both \citep{Moore+1996}, while other, in good agreement with the standard hierarchical scenario of structure formation, could be the result of the merger of smaller disc systems (note that in both cases the ancestors must always be objects of late type).
 
Past efforts to understand S0 and their origin have focused essentially on studying the gross properties of these objects inferred from photometric data in one or more broad wavelength bands, from radio to X-rays. There have been works devoted to analyse, for instance, either the luminosities and sizes of the bulge component \citep{Solanes+1989, Barway+2007, Mishra+2017}, or the global luminosity function \citep{Burstein+2005}, or the Tully-Fisher relation, both standard \citep{Hinz+2003, Bedregal+2006, Williams+2010, Davis+2016} and baryonic \citep{denHeijer+2015}. Such investigations have often produced mixed results regarding the parentage of S0 galaxies, which have led to the proposal of multiple formation theories \citep[see, e.g.][]{vdB2009} whose imprints in the observable properties of the remnants are, nevertheless, difficult to discriminate.

However, it is quite rare to find statistical studies focusing on the stellar content of S0 that deal with the wealth of information that is revealed when the emitted light is divided into its component wavelengths. The main reason is the much longer time-scales required to get a single usable data element, i.e.\ a spectrum pixel with a sufficiently high signal-to-noise ratio ($S/N$), compared to photometric observations which measure light over much broader wavelength bands. Fortunately, this situation has gradually changed over the course of this century thanks to the advent of multi-object spectrographs that allow multiple galaxies to be observed in a single exposure, increasing the census of spectra available to the astronomical community by $10$--$50$ times. One of the most paradigmatic examples of these groundbreaking surveys is perhaps the Sloan Digital Sky Survey \citetext{SDSS; \citealt{York+2000}}, which over the last seventeen years has obtained optical single-fibre spectra for nearly a million galaxies in approximately 8000 square degrees of the sky as regards only the nearby universe ($z\lesssim 0.25$).

This is precisely the approach taken by \citet{Xiao+2016}, who have investigated the nuclear activities of a sample of 583 nearby S0 galaxies observed with the SDSS spectrograph with high $S/N$. After dividing their dataset into galaxies with and without significant nebular emissions, these authors used the BPT diagnostic diagram \citep{Baldwin+1981} which determines the ionisation mechanism from the flux ratios of four main emission lines, to classify their emission-line objects into star-forming galaxies, composite galaxies, Seyfert galaxies, and LINER. Like previous studies have found, their analysis revealed that star-forming S0 have lower stellar masses and luminosities, and bluer global colours than the rest, while a more detailed analysis based on photometric 2D bulge-disc decomposition of a subset of their data reveals that these galaxies also have central regions with a low ($n < 2$) S\'ersic index that, in a substantial number of cases, are bluer than the discs. These results, which agree with photometric studies that also hinted at the existence of two sub-classes of S0 \citep[e.g.][]{Barway+2013}, support a picture in which star-forming S0 form through the accretion of gas from an external source -- a gas-rich dwarf or a cosmological cold-gas filament -- into a progenitor disc galaxy that leads to a short-lived burst of final star formation\footnote{Formation driven by dissipative processes may also explain the ubiquity of S0 with counter-rotating gas and stars \citep[see, e.g.][and references therein]{Tapia+2017}.}. Besides, \citeauthor{Xiao+2016} also find evidences that activity and environment are related, since they observe that S0 with obvious signs of star formation and/or of hosting active galactic nuclei (AGN) reside essentially in low-density regions, while S0 with low-level emission lines ($S/N < 3$) or with classic absorption-line spectra populate environments with a broader range of densities. In an earlier work, \citet{Helmboldt+2008} reached similar conclusions regarding the typical masses and local environments of actively star-forming early-type galaxies.

The spectral analyses of galaxies have recently broken newer ground with the measurement of spectra at multiple points in the same object using fibre bundles closely packed into hexagonally-shaped integral field units. The Mapping of Nearby GAlaxies survey \citetext{MaNGA; \citealt{Bundy+2015}} is a large-scale integral field spectroscopy (IFS) galaxy census included in the ongoing core projects of the fourth SDSS phase \citep{Blanton+2017} that is using the two BOSS spectrographs to provide spatially-resolved spectral information on thousands of nearby objects. In a recent work, \citet{mckelvie+2018} have investigated the stellar populations of both the bulge and disc components of a subset of 279 S0 galaxies selected from a partial release of MaNGA's observations. They extract four Lick indices from the spectral datacubes and use them to measure light-weighted stellar ages, metallicities, and $\alpha$-enhancement parameters. When represented graphically, the index--index diagrams and the implied stellar age--metallicity plots for both regions show bimodal distributions strongly correlated with the stellar mass. The clear bimodality revealed by the spectroscopic data leads these authors to conclude that they are observing two distinct populations of S0 galaxies: one that is old, massive and metal-rich, and that possesses bulges that are predominantly older than their discs, and the other that comprises a younger stellar population, that is less massive and more metal-poor, and that has bulges with more recent star formation than the discs. In agreement with \citet{Xiao+2016}, the correlations inferred are found to extend to the light profiles and colours of the central regions, with low-mass galaxies harbouring a larger fraction of what they call pseudo-bulges and being more star forming, while the opposite is true for higher mass galaxies. This duality in the structural relations of the S0 galaxies points, according to the estimates by \citeauthor{mckelvie+2018}, to separate formation sequences that would be independent of the environment: the low-mass S0 would be the outcome of fading spirals, while their higher-mass counterparts would preferentially arise from mergers. Nonetheless, because nearby cluster S0 are underrepresented in the MaNGA's galaxy sample used, it is also emphasised that this conclusion cannot be extrapolated to regions of high galaxy density.

This is the first article in a series dedicated to investigate the physical properties of the galaxies classified as S0 within the local universe ($z\lesssim 0.1$). Our aim is increasing our understanding of the members of this special morphological class that connects the two extremes of the Hubble sequence and, ultimately, identifying the formation channels that they may have followed. In Section \ref{S:dataset} we give the details of the different databases adopted for this study, which takes as its core data the several tens of thousands optical integrated spectra retrieved from the Main Galaxy Sample (MGS) of the Sloan Legacy Survey \citep{Strauss+2002}, a magnitude-limited spectroscopic catalogue that was completed during the original SDSS observing plan, which ran from $2000$ to $2008$. The single-fibre SDSS spectra are cross-matched with several public photometric datasets, including the recently published catalogue by \citet{Dom&Sanch+2018} of morphologies for MGS galaxies inferred using Deep Learning algorithms and our own determination of the galaxy local density. The adequate exploitation of this large body of data demands the use of automated diagnostic tools capable of extracting as much objective information as possible on the properties of the galaxies. Besides, we want to take advantage of all the information contained in the full range of wavelengths covered by the spectra instead of using just a few spectral lines. To satisfy both requirements, we apply in Section \ref{S:pca} the well-known exploratory technique of Principal Component Analysis \citetext{PCA; e.g. \citealt{Ronen+1999}, and references therein}, which provides an optimal representation of the data in terms of a few mutually-orthogonal linear variables that discriminate most effectively among the galaxy spectra. After discussing the outcome of the PCA in Section~\ref{S:s0_es}, we carry in Section~\ref{S:pc1_pc2} a detailed analysis of the physical characteristics of the major spectral classes into which the S0 population can be subdivided within the subspace defined by the projection of their spectra on the first and second principal components. Finally, in Section \ref{S:summary} we summarise this work, discuss its main findings and present our conclusions. Three Appendices provide evidence of the statistical soundness of our results and complete the information given in the main text.

\section{The Database}
\label{S:dataset}

\subsection{Target selection}
\label{S:targets}

The source of galaxy spectra is the version of the MGS database included in the twelve Data Release of the SDSS \citetext{SDSS-DR12; \citealt{Alam+2015}}. The MGS provides reduced spectroscopic observations for extended objects (galaxies) that are essentially complete for extinction corrected $r$-band Petrosian magnitudes below $r_{\mathrm{lim}}\simeq 17.7$. MGS spectra have been collected from two digital detectors (blue and red) mounted in the same telescope. They consist of 3800 spectral bins which cover a joint wavelength range of $3800$--$9200$ \AA\ with a spectral resolution $R$ ranging from $1850$ to $2200$. This gives an average instrumental dispersion of $69$ km s$^{-1}$ per pixel and a velocity resolution of $\sim 90$ km s$^{-1}$. The aperture diameter of the fibres collecting the spectra is $3$ arcsec. This is a generally small radius within nearby galaxies, meaning that aperture effects could take on certain relevance, especially for the nearest objects. However, as it is empirically demonstrated in Appendix~\ref{sapp:aperture}, the main conclusions derived from this work will not be affected by this potential shortcoming. The global incompleteness of the MGS is small, about 6 per cent \citep{Strauss+2002}, so no action will be taken to compensate it. Similarly, we will ignore the additional incompleteness that arises in SDSS spectroscopy at very low redshifts ($z < 0.05$) for very bright galaxies due to blending with saturated stars, as it is negligible, less than $0.5$ per cent according to \citeauthor{Strauss+2002}.

Galaxy morphologies are retrieved from the catalogue by \citet{Dom&Sanch+2018} that lists this information for $\sim 671{,}000$ galaxies from the SDSS-DR7, covering most of the MGS. The classification of such a huge number of objects is obtained thanks to Deep Learning algorithms that use Convolutional Neural Networks trained with colour images from both the Galaxy Zoo 2 \citep{Willett+2013} and the visual classification catalogue of \citet{Nair&Abraham2010}. Compared to traditional visual determinations of the Hubble type, this automated classification is more reliable, more objective, and provides a more accurate identification of the structural elements needed to distinguish among the different galaxy classes. Thanks to these characteristics the morphological identifications extend to up to the limits of the MGS and, therefore, are applied to a very large number of objects covering a wide dynamical range of physical properties.

\citeauthor{Dom&Sanch+2018}'s morphologies are based on the training with the  \citeauthor{Nair&Abraham2010}'s dataset. They are provided in terms of a continuous numerical parameter which we convert into the classical discrete $T$-type classification (see e.g. \citealt{Vaucouleurs+1977} and references therein) by simply taking the nearest integer to the reported values. According to these authors, the average scatter of Deep Learning morphological classifications is only $\sigma = 1.1$ and their performance, in terms of accuracy, completeness, and contamination, comparable or even better than that of expert classifier intercomparisons. To efficiently distinguish between pure E and S0 galaxies, they also provide the probability of being S0 ($P_{\mathrm{S0}}$) for objects with $T\leq 0$. We will consider any galaxy with $T \leq 0$ and $P_{\mathrm{S0}} > 0.7$ to be an S0. Such a conservative value for the probability has been chosen to prioritise the purity of the S0 sample while preserving a sufficient number of objects for the analysis. Possible contamination by the roundest E, which are difficult to distinguish from relatively face-on S0, or by undetected spiral structure in highly inclined disks is in all likelihood minimal thanks to the strict selection criterion adopted and, especially, due to the fact that computers are far better than humans at detecting subtle variations in the brightness of the images \citetext{see also the statements by \citealt{Fischer+2019} as regards the identification of S0 classified using the same methodology}. Visual inspection of the images of a random subset of our data has reassured us that we are dealing with genuine representatives of this population. We provide some examples in Appendix~\ref{app:spectra}.

Although the spectroscopic MGS observations span a relatively broad range of redshifts, we deal only with those S0 galaxies lying within $0.01\le z \le 0.1$. The upper limit, which encompasses the peak of the MGS, is set to select a homogeneous sample of lenticular objects representative of the local universe over which the effects of K-corrections, curvature, and cosmic evolution are negligible. The lower redshift limit excludes those (few) galaxies with most uncertain measurements due to their closeness or large apparent brightness. Although their number is also very small, we have also excluded galaxies with apparent magnitudes from elliptical Petrosian $r$-band fluxes brighter than $12.0$ to eliminate objects with unrealistically high absolute luminosities in the surveyed volume. The application of all these constraints allows us to obtain a magnitude-limited sample of $68{,}043$ integrated spectra from S0 galaxies that will be hereafter referred to as the Main Local Sample of S0 (MLSS0).

The upper redshift adopted for the MLSS0 has also the characteristic of maximising the number of MGS galaxies that can be included in a volume-limited subset of it. By taking into account the redshift and magnitude limitations commented above, we define a volume-limited sample of $32{,}188$ S0, the VLSS0 hereafter, that includes galaxies with Petrosian $r$-band absolute magnitude $M_r \lesssim -20.5$. Note that the sizes of the two collections of galaxies that we have just defined are approximately two orders of magnitude larger than that of any set of integrated S0 spectra used in previous studies and have no parallel either in the scale of S0 samples recently extracted from IFS surveys, which, at the time of the writing of the present work, are still insufficient to draw conclusions with a high statistical power.

\subsection{Spectrophotometric data}
\label{S:properties}

A great deal of the measurable properties required for this and forthcoming studies are selected mainly from the v1\_0\_1 of the NASA-Sloan Atlas (NSA) catalogue \citep{Blanton+2011}. The NSA dataset includes virtually all galaxies with known redshifts out to about $z < 0.15$ within the coverage of the SDSS-DR11 (an internal bookkeeping release). This catalogue is built around the SDSS optical/NIR properties, which are completed with observational data from the ultraviolet Galaxy Evolution Explorer (GALEX) survey \citep{Boselli+2011}, the $21$-cm Arecibo Legacy Fast ALFA (ALFALFA) survey \citep{Giovanelli+2005} -- of which we also have the 21-cm line widths and masses of neutral hydrogen (HI) of its final version \citep{Haynes+2018} --, as well as with some measurements from the CfA Redshift Catalogue \citetext{ZCAT; \citealt{Huchra+1995}}, the 2dF Galaxy Redshift Survey \citetext{2dFGRS; \citealt{Colless+2001}}, the 6dF Galaxy Survey \citetext{6dFGS; \citealt{Jones+2004}}, and the NASA/IPAC Extragalactic Database (NED). However, not all the data in the NSA catalogue have been used. Some information, such as its estimation of the stellar mass, has been replaced with measurements retrieved from the latest version of the GALEX-SDSS-WISE Legacy Catalogue \citetext{GSWLC-2; \citealt{Salim+2016, Salim+2018}}, which has led us to recalculate the mass-to-$r$-band-light ratio (unless otherwise stated, we use elliptical Petrosian aperture photometric data). We have also taken from this latter catalogue the estimates of the star formation rate (SFR) which, like the stellar mass, is based on joint UV+optical+mid-IR SED fitting, as well as the values of dust attenuation in different bands that we use to study the effects of internal extinction in our galaxies. In addition, the Portsmouth stellar kinematics and emission-line flux measurements of \citet{Thomas+13} have been chosen as a source for stellar velocities, fluxes and equivalent width measurements of the most important recombination lines from SDSS spectra, whilst mass-weighted stellar population ages and metallicities calculated using a Chabrier-MILES fit of the stellar population parameters are retrieved from the eBOSS Firefly Value-Added Catalogue \citep{Comparat+2017}. All these data have been further augmented with the information about group membership listed in the catalogue by \citet{Tempel+2017}. Finally, for galaxies belonging to the VLSS0, the information on the physical properties has been completed with our own determination of the number density of their environment (see next section). 

Unless otherwise stated, all cosmology-dependent variables involved in the present investigation have been scaled to a standard flat Friedmann-Robertson-Walker world model with matter energy density $\Omega_{\text{m}}=0.3$, dark energy density $\Omega_\Lambda = 0.7$, and Hubble constant $H_0 = 100 h$ km s$^{-1}$ Mpc$^{-1}$ with $h = 0.7$.

\subsection{Local densities}
\label{S:localmu}

The proxy for environment we use is our own estimate of the local density of galaxies within narrow redshift slices. Extinction-corrected galaxy densities have been calculated exclusively for all the members of our volume-limited sample that belong to the Northern Galactic cap of the MGS, since in this region of the sky the sampling of galaxies is more exhaustive than in the Southern Galactic cap.

For a given galaxy, the density of its local environment is computed using an optimised version of the $k$-nearest neighbour Bayesian estimator \citep[cf.][]{Casertano&Hut1985}
\begin{equation}
\mu_5 = C(\Delta m) \cdot \left(R^2 \cdot \sum_{i=1}^5 d_{i}^2\right)^{-1}\;,
\label{eq:density}
\end{equation}
where $R \approx cz/H_0$ is the radial distance to that galaxy and $d_{i}$ is the angular separation between it and its $i$th closest neighbour selected among those galaxies with a recessional velocity within $1000$ km s$^{-1}$ from the target. Note that the use of the distances to all five neighbours notably improves the precision of density estimates compared to the traditional $k$-nearest neighbour metric, which only considers the distance to the farthest companion. Besides, we have also included in Equation~(\ref{eq:density}) a correction factor,
\begin{equation}
C(\Delta m) = A\cdot 10^{0.6\Delta m}\;,
\label{eq:corr_extinc}
\end{equation}
to account for the effect of Galactic extinction on the observed densities \citep{Solanes+1996}, something we consider necessary given the vast area of the sky probed by the sample. In Equation~(\ref{eq:corr_extinc}), $\Delta m$ is the extinction in the $r$-band at the position of the target, which we retrieve from the SDSS photometry, and $A$ is a normalisation constant that is set to one, as we are only interested in ranking the local densities. Note also that by limiting the search of neighbours to galaxies lying within thin shells in recessional velocity around the target galaxy, Equation~(\ref{eq:density}) becomes, in practice, a three-dimensional local density estimator.

We have avoided calculating densities for objects near the edges of the dataset, as they may be underestimated. For this, we have first computed the probability density function of the angular distances to the fifth neighbour of all the Northern Galactic cap members of our volume-limited subset. Then, we have taken three times the scale (that is, the standard deviation) of this distribution to establish the thickness of the band around the edges of the survey in which the density calculation has not been performed -- the galaxies lying in this peripheral region have nonetheless been used in the calculation of the densities of objects lying further inside. Likewise, the edge effects in the density estimates of the nearest and furthest galaxies have been taken care of by increasing by 1000\kms\ the radial coordinate of our dataset at both ends, discarding afterwards the galaxies included in the resulting extra volume. 

\section{Principal Component Analysis}
\label{S:pca} 

Because we are dealing with spectra of galaxies at distances that range over several hundreds of Mpc, their total fluxes and wavelength coverage are not directly comparable. It is therefore necessary to preprocess all the spectra to put them on an equal basis. We firstly start by shifting each spectrum to the laboratory rest-frame. Since the binning in wavelength has a constant logarithmic dispersion, this correction is given by $\log \lambda_{\mathrm r} = \log \lambda_{\mathrm o} - \log (1 + z)$, where $\lambda_{\mathrm r}$ is the rest-frame wavelength and $\lambda_{\mathrm o}$ is the observed one. After applying this correction we re-bin the flux and its error, interpolating into pixels with a constant logarithmic spacing of $0.0001$, so the resolution of the original spectra is preserved along the full visible range\footnote{\url{http://www.sdss.org/dr12/spectro/spectro_basics/}}. 

We have also taken care of spectral bins affected by sky lines or bad data by blacking out those pixels whose errors are set to infinity as well as those that have the mask bit \texttt{BRIGHTSKY} activated\footnote{According to the SDSS spectral mask criteria, \url{http://www.sdss.org/dr12/algorithms/bitmasks/\#SPPIXMASK}.}. Any spectrum containing more than $10$ per cent of troublesome pixels according to this criterion is fully discarded. 

The spectra that pass our quality filter are then normalised. We have followed \citet{Dobos} and rescaled each individual spectrum to have the average value of the flux in several intervals representative of the continuum level, which is set equal to 1. We use the following expression: 
\begin{equation}
	f_{ij}^n = f_{ij} \cdot \left(\sum_k \dfrac{f_{kj}}{N_j}\right)^{-1}\;,
	\label{eq:fnorm}
\end{equation}
where $f_{ij}$ is the flux at spectral bin $i$ of galaxy $j$ and the sum runs over the $N_j$ bins conforming the following four wavelength intervals devoid of strong lines: $4200\:$--$\:4300$, $4600\:$--$\:4800$, $5400\:$--$\:5500$ and $5600\:$--$\:5800$ \AA. Note that $N_j$ may differ from one galaxy to another depending on the number of pixels with bad data that fall on these continuum regions. Spectra with more than $20$ per cent of defective pixels within such intervals are also discarded. 

The removal of problematic pixels introduces null values in the array of normalised fluxes, thus preventing the proper performance of the PCA. To solve this problem, we employ the PCA-based algorithm described by \citet{Yip} to fill such gaps. A brief summary of the algorithm's functioning would be that, on a first stage, it fills the gaps by linear interpolation and then uses a linear combination of the eigenspectra (ES from now on), derived from the full dataset, to substitute the missing data and re-fill the gaps with more realistic flux values \citep{Connolly}. This second step is iterated until the reconstructed fluxes from the empty pixels converge. With this stage completed, the spectra are corrected for Galactic dust reddening calculated by means of the extinction python module \url{https://extinction.readthedocs.io/en/latest/} that relies on the standard \citet{Fitzpatrick1999} dust extinction model. The $A_V$ coefficient per object is obtained from the Galactic extinction in the $g$ band included in the NSA catalogue assuming a fixed $R_V$ of 3.1.

\begin{figure}
	\includegraphics[width=\columnwidth]{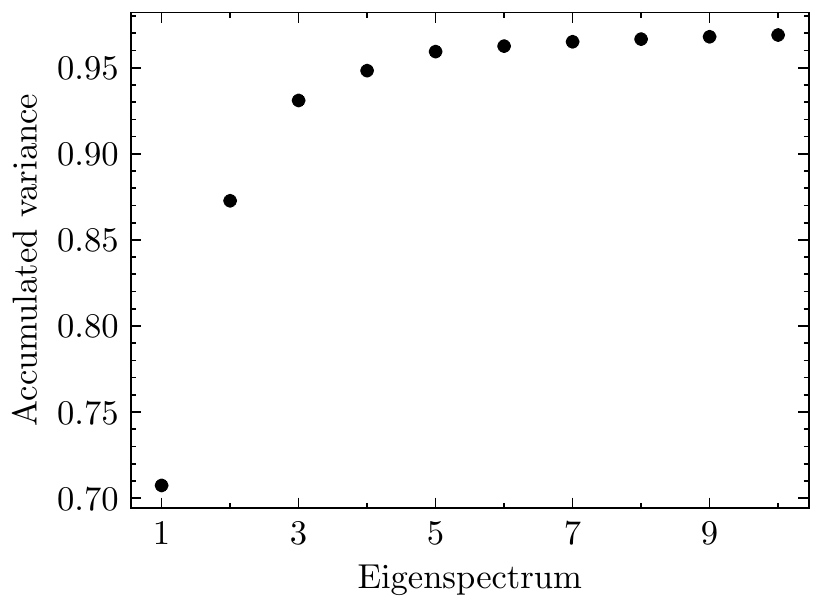}
    \caption{Accumulated fractional variance from the first ten eigenspectra of the training sample.}
    \label{fig:samp_var}
\end{figure}

Following preprocessing, we proceed to define the subset of the MGSS0, formed by the $6477$ S0 galaxies whose spectra have $S/N \geq 30$, as the training sample for the computation of the PCA. This restriction of the original sample to the galaxies with highest $S/N$ is adopted to guarantee that the different dimensions inferred from the PCA decomposition are entirely physically motivated and not affected by noisy data. On the other hand, the threshold applied in the $S/N$ maximises the amount of variance explained by the different components, i.e.\ optimises the dimensionality reduction, allowing us to work with the smallest possible set of ES, while still dealing with a training subset sufficiently large to ensure that these eigenvectors, also called Principal Components, are representative of the whole dataset. As shown in Fig.~\ref{fig:samp_var}, the first three ES already account for more than $90$ per cent of the variance of the training sample, whilst only four ES are required to achieve the $95$ per cent of it (had we derived the ES using the whole volume-limited sample, the number of dimensions required to account for these fractions of the variance would have been substantially larger). Besides, we have verified that the ES from this training sample are capable of reconstructing the specific spectrum of any S0 that belongs to the parent MGSS0 dataset. This can be seen in the different panels of Fig.~\ref{fig:spec_samp2} in Appendix~\ref{app:spectra}, where we show, using arbitrary examples of S0 spectra, that the linear combination of the mean spectrum (top panel of Fig.~\ref{fig:ES}) with the first ten ES -- the latter weighted by the coefficients that result from projecting the individual spectra on the corresponding ES -- gives rise to reconstructed spectra nearly indistinguishable from the input ones. These projections of the spectral data vectors on the new orthogonal basis defined by the Principal Components, are hereafter labelled PC1, PC2, etc. 

\section{S0 Eigenspectra}
\label{S:s0_es}

The results of the previous section tell us that it should be possible to obtain a good insight on the main physical characteristics of the S0 population in the local universe by plotting their spectral data projected on a low-dimensional space whose orthogonal axes are formed by the first three ES. 

\begin{figure*}
	\includegraphics[width=\textwidth]{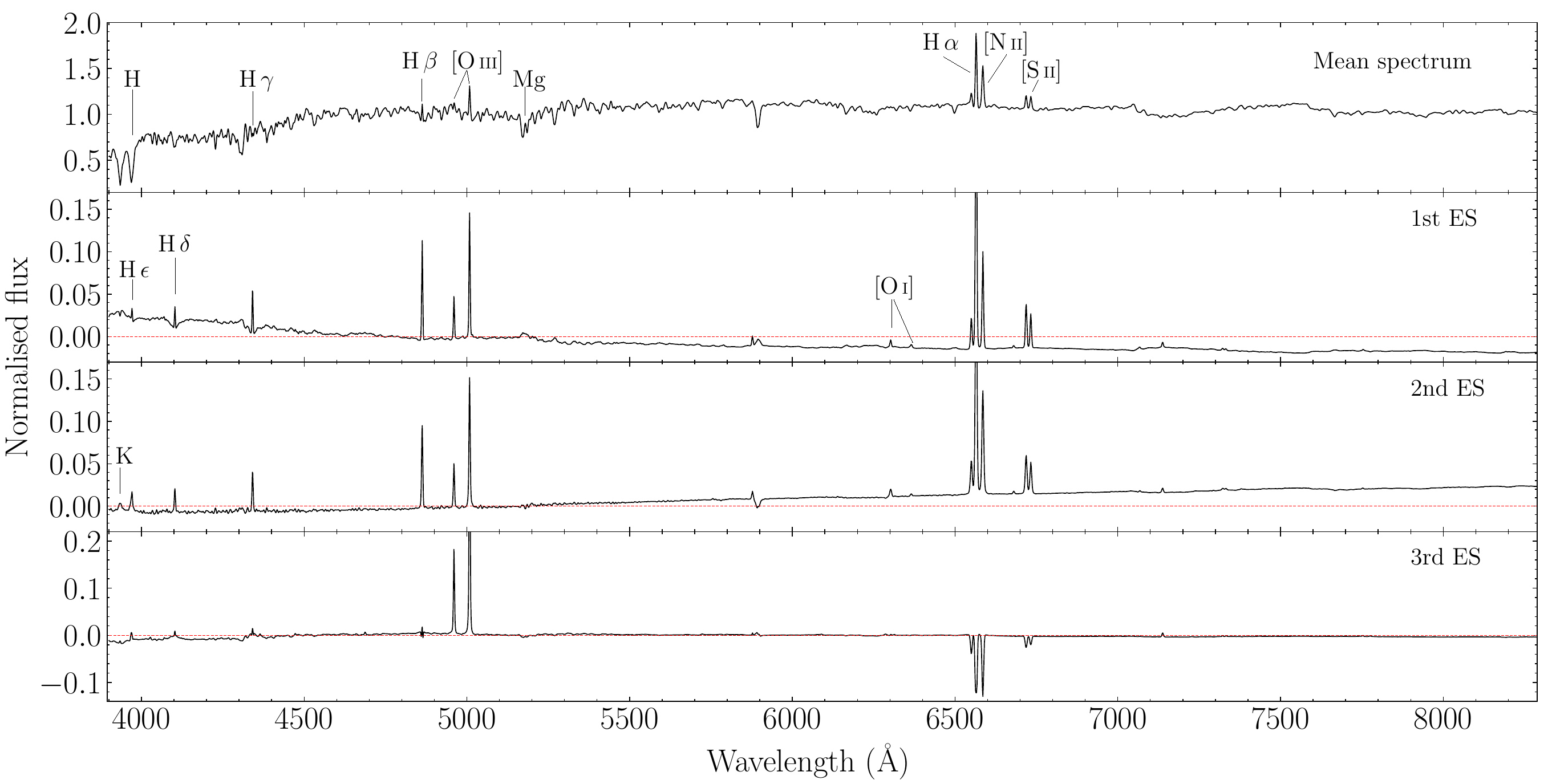}
\caption{From top to bottom, mean spectrum and first three ES of the $6477$ S0 galaxies included in the training sample. The zero level of the ES is highlighted with a horizontal red dashed line. Some of the most important lines of the mean spectrum, also visible in the other panels, are labelled. Lines such as [\ion{O}{I}], $\text{H}\,\delta$ and $\text{H}\,\epsilon$, which is blended with the \ion{Ca}{II} H line, appear in emission in the first two ES, but not in the mean spectrum, whilst the other member of the \ion{Ca}{II} doublet, the K line, is seen, also in emission, only in the second ES. Note, however, that the true emission/absorption nature of these lines is determined in practice by the sign of the weight factor multiplying each ES. The variance of the training sample explained by the $1$st, $2$nd and $3$rd ES is $71$, $17$ and $6$ per cent, respectively, in all a total of $93$ per cent.}
    \label{fig:ES}
\end{figure*}

Fig.~\ref{fig:ES} depicts the mean of the $\sim 6500$ training spectra along with the first three ES. The typical S0 spectrum, depicted in the top panel, is basically characterised by a red continuum that in the violet part shows a relatively strong $4000$ \AA\ break ($D4000$) and conspicuous H and K absorption lines corresponding to the fine structure splitting of the singly ionised calcium (\ion{Ca}{II}) at $\lambda =3935$ \AA\ (K) and $3970$ \AA\ (H)\footnote{Rest-frame wavelength values are expressed in vacuum throughout the paper.}. The average spectrum also shows a moderately strong $\text{H}\,\alpha$ emission line at $6565$ \AA\ and other, weaker, nebular lines that are labelled in the figure. The second panel from the top shows the first ES, which accounts for 71 per cent of the variation in the training sample. This mode combines a blue continuum with particularly strong emission lines that include the Balmer series from $\text{H}\,\alpha$ to $\text{H}\,\epsilon$, as well as forbidden emission lines from both low-ionisation species such as [\ion{S}{II}], [\ion{N}{II}], and [\ion{O}{I}], and one, [\ion{O}{III}], that is highly ionised (bear in mind, however, that the lines in the ES may act in practice either as emission lines or as absorption lines, depending on the sign of the weight factor applied to them). The second ES (third panel from the top), which accounts for 17 per cent of the variation in the training sample, has a similar appearance to the former spectra in terms of the emission lines. However, it differs basically in two aspects, the most important being that in this ES the continuum is red, whilst the other difference concerns the presence of the \ion{Ca}{II} doublet in emission. It is also evident that the two components of this doublet behave somewhat differently, almost certainly due to the close proximity between the H line of Calcium and the Balmer $\text{H}\,\epsilon$ line, whose strengths are also often anti-correlated. This second component may play a role in the regulation of the strength of the emission and absorption lines present in the average spectrum. The third ES, shown in the bottom panel, accounts for only 6 per cent of the total variation, meaning that its features provide only a relatively fine tuning to the spectra reconstructed from the two lower order ES. In this mode the continuum plays virtually no role -- it is basically null except at wavelengths $\lesssim 4500$ \AA, where it slightly drops towards negative flux values --, the most important characteristics being two strong [\ion{O}{III}] emission lines, at $4960$ and $5008$ \AA, and the $\text{H}\,\alpha$+[\ion{N}{II}]+[\ion{S}{II}] complex, which is now observed in absorption. Other lines such as [\ion{O}{I}], higher-order Balmer lines, and the H line of \ion{Ca}{II} are also present, although they are all very weak. As in the case of $\text{H}\,\alpha$ and the adjacent [\ion{N}{II}] and [\ion{S}{II}] lines, the [\ion{O}{III}]$\lambda 5008$ emission line is not just a star-forming indicator but also a good tracer of AGN activity \citep[see e.g.][and references therein]{Trouille+Barger2010,Suzuki+2016}. This is because although this forbidden emission line originates from high ionisation states caused by both hot, young massive stars and AGN, it has been observed to be relatively weak in metal-rich, star-forming galaxies. Indeed, thanks to these characteristics the third principal component is, despite its relative low weight, a useful diagnostic tool for activity in S0 galaxies (Tous et al. 2020, in preparation). 

\begin{figure}
	\includegraphics[width=\columnwidth]{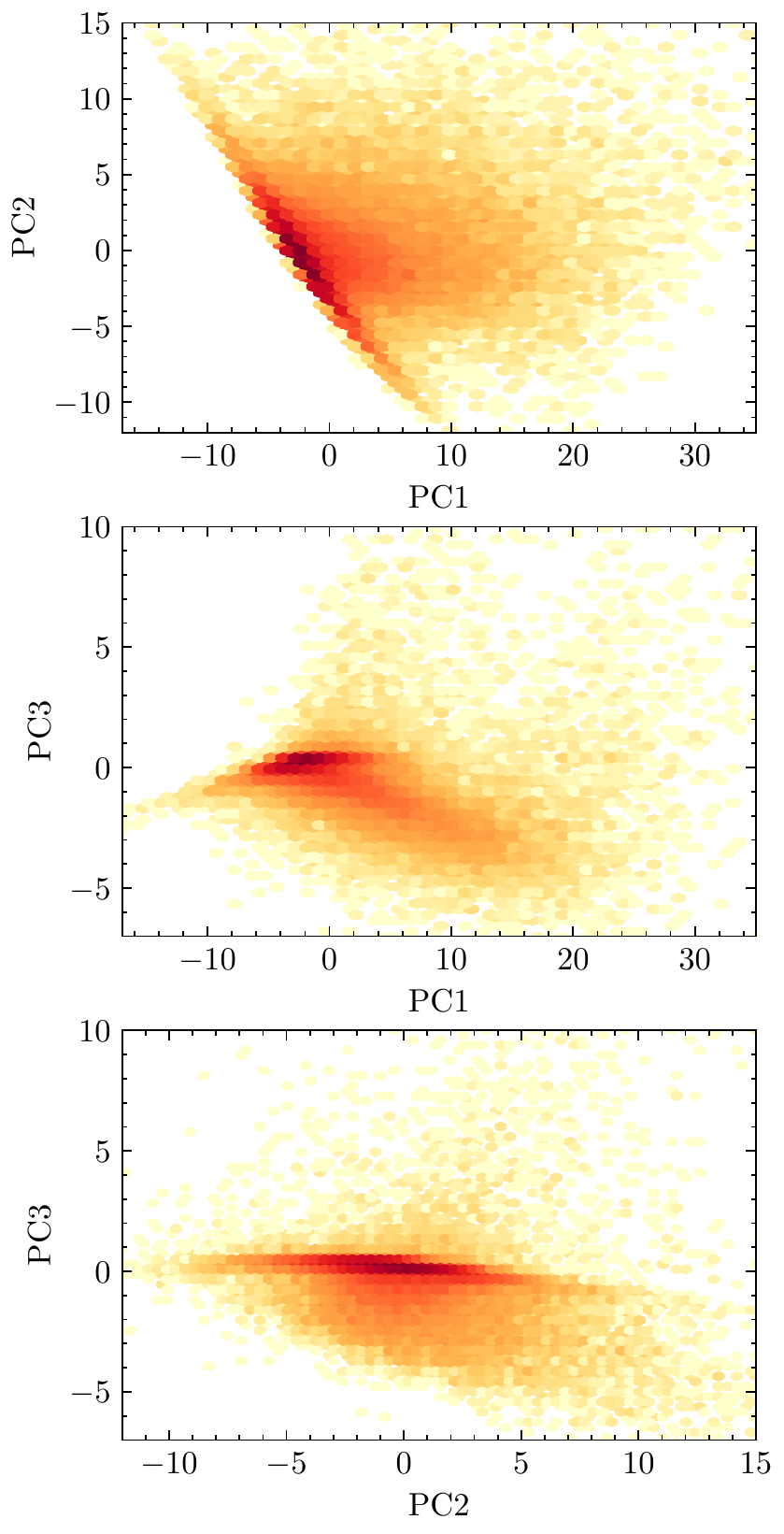}
    \caption{Projections of the S0 spectra in the MLSS0 on the three planes defined by the first three Principal Components. Data points have been grouped in hexagonal bins where the intensity of the colour scales with the logarithm of their number density. From top to bottom the panels portray the PC1--PC2, PC1--PC3, and PC2--PC3 subspaces.}
    \label{fig:pcx_vs_pcy}
\end{figure}

In Fig.~\ref{fig:pcx_vs_pcy}, we portray the pairwise scatter plots that result from plotting the first three PC of the S0 galaxies in the MLSS0 against each other. As shown in the top panel, the PCA results delineate three distinct areas in the PC1--PC2 subspace, which point to the existence of two different sub-populations (from a spectral point of view) of S0 galaxies. A large number of points appear concentrated in a crowded narrow band that crosses the diagram diagonally defining a zone in this plane in which the values of the first two transformed predictors show a strong linear correlation indeed. This tilted band has a sharply-defined edge on its left side setting the boundary of a forbidden region in this subspace (see Section~\ref{S:pc1_pc2}), whilst it extends on the right into a substantially less populated and much more scattered cloud of points where the projected values conform to the expectation of being linearly uncorrelated. The compartmentalisation of space observed in the PC1--PC2 diagram is not preserved in the other two 2D subspaces represented in the mid and bottom panels of Fig.~\ref{fig:pcx_vs_pcy} that involve the projections on to the third principal component. The distributions of data points in the PC1--PC3 and PC2--PC3 diagrams are certainly reminiscent of that in the top panel, but show a lower degree of organisation. It can be seen that the galaxies whose spectra delineate the diagonal strip observed in the PC1--PC2 subspace are still strongly clustered, now forming in both planes an approximately horizontal strip around the PC3$\;= 0$ coordinate. The rest of the data distribute unevenly both above and below this band, with the largest fraction showing negative values of PC3 and, at the same time, values of the PC1 and PC2 predictors that are positive in a good measure, an inherent feature of galaxies with a high $\text{H}\,\alpha$/[\ion{O}{III}] flux ratio. Whichever the case, in the last two subspaces the boundary of the forbidden region is much less clearly defined. Also note the fact that the peaks of the point clouds of the three 2D projections fall near the origin  of the coordinate system, which indicates that the spectra of a good number of S0 galaxies must be quite similar to the mean spectrum reported in Fig.~\ref{fig:ES}.

\section{The PC1--PC2 subspace}
\label{S:pc1_pc2}

In the remainder of the article, we will focus mainly on the definition and analysis of the major spectral classes into which the S0 population can be subdivided within the PC1--PC2 subspace, while the projections of the S0 spectra on the third principal component that, as mentioned in the previous section, bear information on the star formation rates and nuclear activity of these galaxies, will be investigated in future work. For simplicity, in this and the following sections the discussions involving the principal components will be dealt with using their observed values, since the correction of the aperture and inclination biases outlined in Appendix~\ref{app:robustness} only lead to small differences in the results that do not alter the conclusions of this work in a significant way.

\subsection{Relationship with spectrophotometric observables}
\label{S:proxy}

We have just seen that a great deal of the information about the variance of the S0 spectra is contained on the first two eigenvectors, which are dominated by strong emission lines but show continua with slopes of opposite sign (see Fig.~\ref{fig:ES}). These two ES when combined with the mean spectrum must reproduce the general spectral features characteristic of completely passive galaxies with no detectable emission lines (this requires both ES to have negative weight factors; see e.g.\ panel c in Fig.~\ref{fig:spec_samp2}), as well as of galaxies showing star formation/starburst/AGN signatures, either on top of a red or blue continuum (the latter requires PC1$\;>0$, or PC2$\;<0$, or both; see e.g.\ panels (h), (i) and (k) of Fig.~\ref{fig:spec_samp2}). Thus, although there is no guarantee a priori that the new dimensions provided by a PCA -- they are mutually orthogonal linear combinations of the input variables -- have any physical meaning, it is sensible to expect in this particular case that the main spectral predictors we have inferred are closely connected to parameters that inform about the star formation history of the galaxies. 

\begin{figure}
\includegraphics[width=\columnwidth]{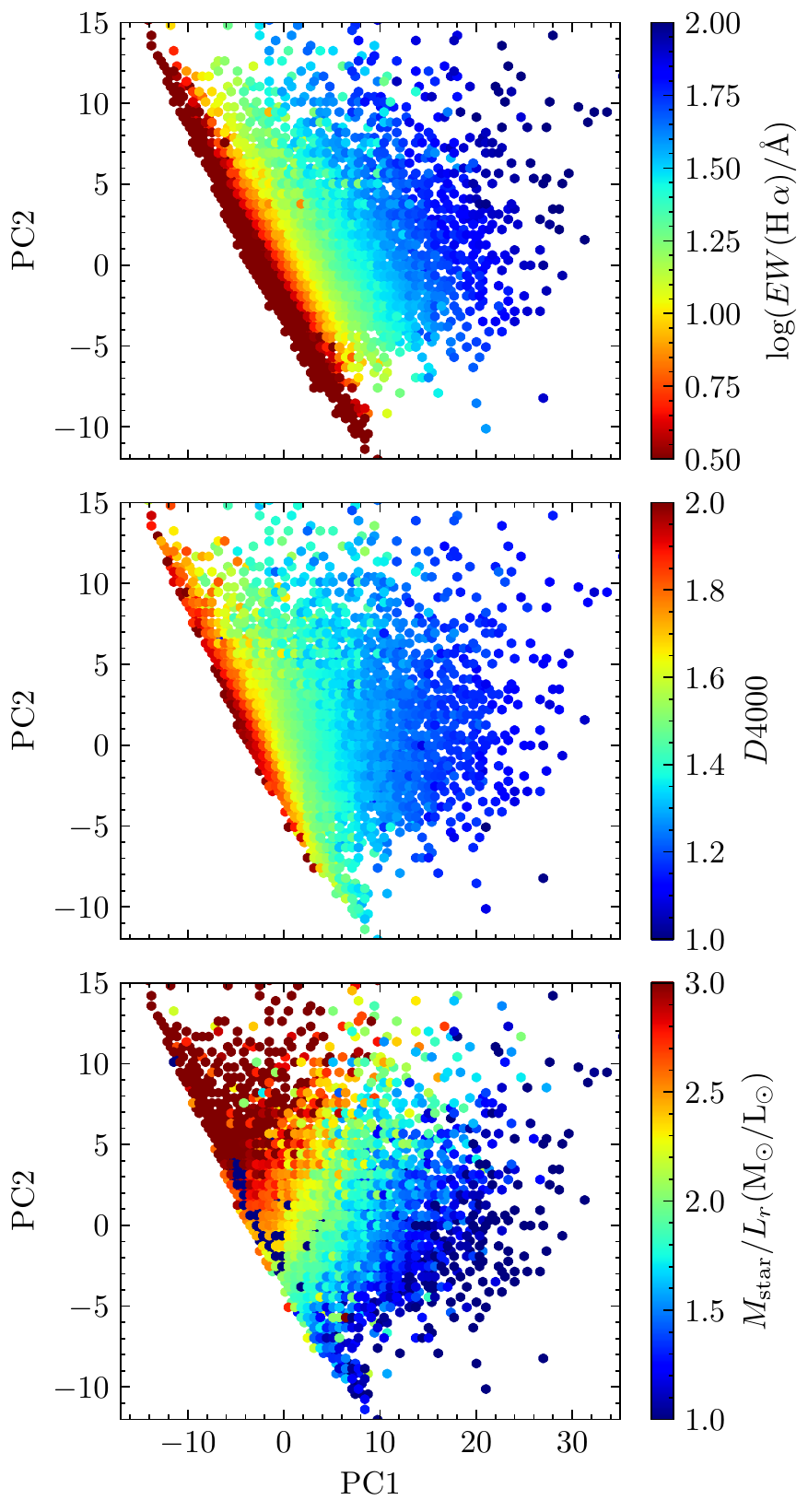}
\caption{Same representation as in the upper panel of Fig~\ref{fig:pcx_vs_pcy} but using colour as a third dimension to indicate the mean equivalent width of the $\text{H}\,\alpha$ line at each hexagonal bin (top panel), the mean $D4000$ break (mid panel), and the mean stellar mass-to-light ratio $M_\text{star}/L_r$ (bottom panel).}    
\label{fig:pc1_pc2_prop}    
\end{figure}

\begin{figure*}
	\includegraphics[width=0.8\textwidth]{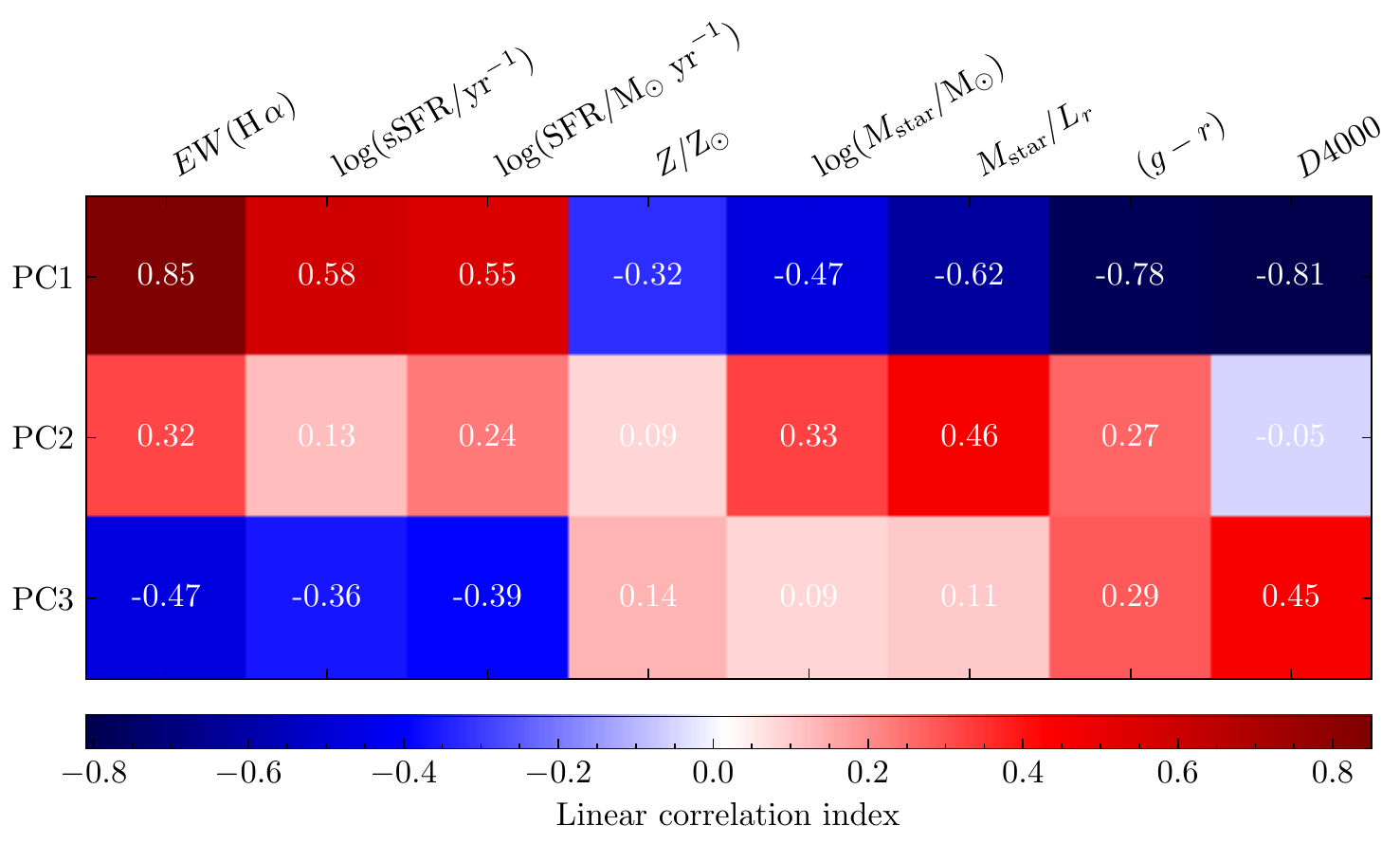}
\caption{Linear pairwise correlations between the first three principal components of the spectra of S0 galaxies and the eight spectrophotometric properties in our database more closely related to them. Cells are colour-coded according to the values of the Pearson's correlation statistic calculated from the VLSS0, which are also annotated.}    
\label{fig:CorrMat}    
\end{figure*}

Thorough review of the different spectrophotometric properties included in our final database has shown that the direction of the maximum variance in the strength of the $\text{H}\,\alpha$ emission line, given by its equivalent width, $\text{EW}(\text{H}\,\alpha)$, is essentially contained in the PC1--PC2 subspace. Besides, it is nearly perfectly orthogonal to the direction delineated by the tilted band identified in the top panel of Fig.~\ref{fig:pcx_vs_pcy}, as illustrated in the top panel of Fig.~\ref{fig:pc1_pc2_prop}. Actually, the $\text{EW}(\text{H}\,\alpha)$ is the parameter of our catalogue that shows the strongest positive linear correlation with the first PC, whilst it also shows a weak positive correlation with PC2: according to the annotated heatmap depicted in Fig.~\ref{fig:CorrMat} the corresponding Pearson correlation coefficients are $r[\text{EW}(\text{H}\,\alpha),\text{PC1}]=0.85$ and $r[\text{EW}(\text{H}\,\alpha),\text{PC2}]=0.32$, respectively. Besides, Fig.~\ref{fig:pc1_pc2_prop} allows one to see that the sharp lower-left boundary of the data cloud is the locus of S0 galaxies with no emission in the $\text{H}\,\alpha$ line. This sensibility of the PC1--PC2 subspace to a specific spectral line may come as a surprise as PCA is known not to do a good job detecting differences in narrow spectral features. However, the fact that the $\text{H}\,\alpha$ line is the strongest emission line in the mean spectrum and that it is flanked by two [\ion{N}{II}] lines that are also rather strong (see Fig.~\ref{fig:ES}) make the $\text{H}\,\alpha$--[\ion{N}{II}] complex a remarkable classification feature within any optical galaxy spectrum. Nor must we forget that the $\text{H}\,\alpha$ emission can be closely related to quantities that are sensitive to the past and present star formation of the galaxies, such as the $D4000$ break, integrated colours, and SFR, both global and per unit mass, that involve more significant portions of the flux contained within the optical window. Indeed, we find that these latter parameters are also contained in the PC1--PC2 plane in good measure (see e.g.\ the mid panel of Fig.~\ref{fig:pc1_pc2_prop}) and, similarly to the $\text{EW}(\text{H}\,\alpha)$, they show a moderate to strong correlation with PC1 and a substantially feeble relationship with PC2. Actually, the correlations between all the physical properties investigated and the second predictor -- which, in general, are positive -- are found to be weak at best (all Pearson's coefficients are $< 0.5$). As shown in Fig.~\ref{fig:CorrMat}, the stellar mass-to-light ratio $M_\text{star}/L_r$ whose variance, like that of the colour -- with which is known to be closely related -- is also largely encapsulated into the PC1 axis (see the bottom panel of Fig.~\ref{fig:pc1_pc2_prop}), is the parameter that shows the most evident connection with PC2 ($r[M_\text{star}/L_r,\text{PC2}]=0.46$). For completeness, we have also included in Fig.~\ref{fig:CorrMat} measures of the strength of the linear association between the physical variables and PC3. They behave much like those involving PC1, but are somewhat weaker (the strengths are similar to those involving PC2) and have opposite signs. All these results confirm that the most relevant predictors derived from our characterisation of the S0 spectra encode information about the secular evolution of the stellar population of these galaxies.

\subsection{Main spectral classes}
\label{S:split}

As we have already pointed out, the cloud of S0 points in the PC1--PC2 plane shows obvious signs of being distributed into two regions (Fig.~\ref{fig:pcx_vs_pcy}), whose natural divider appears to be a straight line orthogonal to the direction that maximises the variation of the $\text{EW}(\text{H}\,\alpha)$ (see the top panel of Fig.~\ref{fig:pc1_pc2_prop}). However, although the existence of a thin, lens-shaped zone consisting essentially of the S0 galaxies without significant $\text{H}\,\alpha$ emission is undeniable, it is not at all obvious to find the best place where to put the divider between said narrow band and the much more extended and scattered region dominated by the S0 with emission lines. Thus, rather than attempting to split the data by eye, and in line with the general approach adopted in this work, we have considered more adequate using a supervised machine learning classifier to split the distribution of S0 data points in this subspace into two different classes. 

\begin{figure*}
    \centering
	\includegraphics[width=1.1\columnwidth]{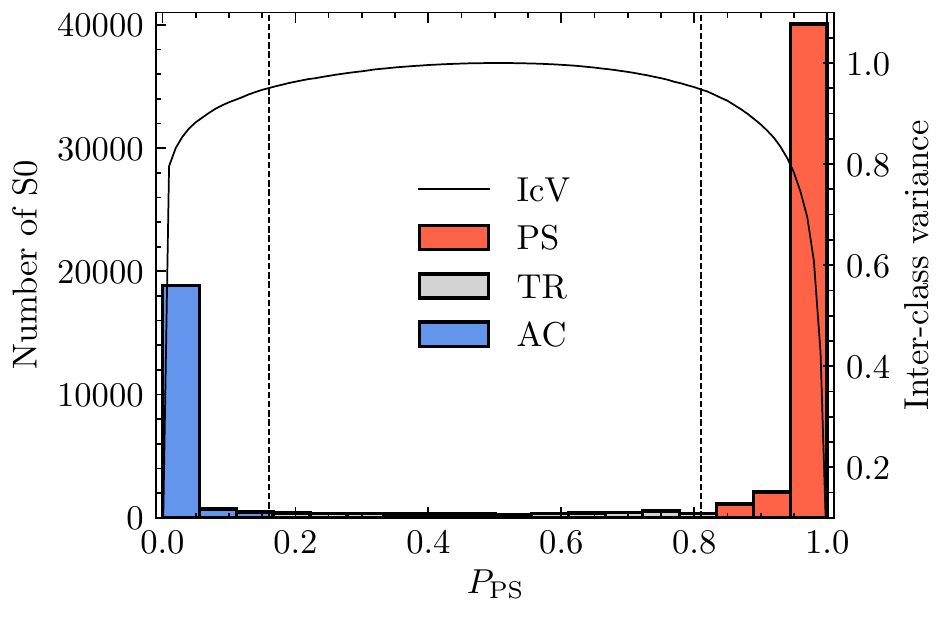}\hfill
	\includegraphics[width=0.95\columnwidth]{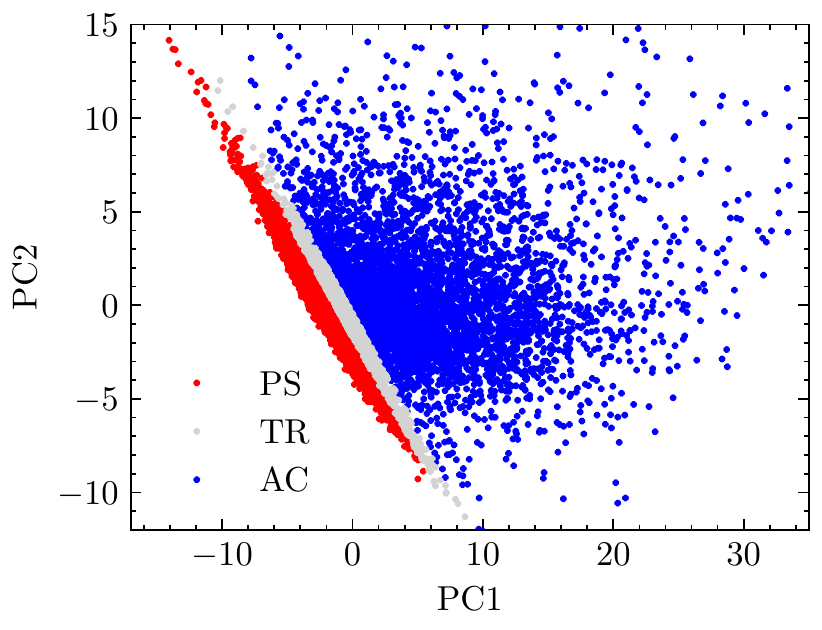}
    \caption{\emph{Left}: Histogram showing the probability distribution that an S0 belongs to the Passive Sequence, $P_{\mathrm{PS}}$, which results from the application of the Logistic Regression to the whole MLSS0 in the PC1--PC2 subspace. The black curve indicates the fractional inter-class variance (IcV). We choose a value of $0.95$ of this parameter to split the original dataset into PS objects (S0 with $P_{\mathrm{PS}} \geq 0.81$), shown in red, and  Active Cloud (AC) members (S0 with $P_{\mathrm{PS}} < 0.16$), which are shown in blue. S0 galaxies with values of $P_{\mathrm{PS}}$ outside these two extreme ranges, identified in the plot by two vertical dashed lines, are assigned to the Transition Region (TR) and are expected to show spectral characteristics intermediate between those of the PS and AC objects. \emph{Right}: similar to the top panel Fig.~\ref{fig:pcx_vs_pcy} but, instead of binning the data points into hexagons, here we represent a random $20\%$ of the original sample (to avoid overcrowding) with the different features of the PC1--PC2 diagram highlighted in colour: PS (red), AC (blue), and TR (grey).}
    \label{fig:split}
\end{figure*}

After exploring several binary classification algorithms, we have found that, for the sort of distribution we are dealing with, a Logistic Regression offers the best performance as it yields the highest accuracy, which is evaluated as the ratio of correctly classified objects over the total members of the training sample. The classification has been directly implemented in the PC1--PC2 plane and not in a subspace of higher dimension. Apart from a matter of simplicity, we have proceeded in this way mainly because we have verified that the outcome of the process is insensitive to the amount of PC involved. We have employed as training dataset a sample of $400$ S0 spectra selected from bona-fide members of the two main groups we want the algorithm to come up with, tagging the galaxies falling into the narrow, lens-shaped region with ones and with zeros those outside it. Basically, what the algorithm does is first to fit a logistic model to the training set according to the labels assigned to each point and the positions they occupy in the PC1--PC2 subspace. Then, the model is used to assign to each galaxy in the whole dataset a probability of belonging to the most compact mode, $P_{\mathrm{PS}}$. The outcome of this procedure is the extreme bimodal distribution shown in the histogram depicted in the left panel of Fig.~\ref{fig:split}, where the data are essentially grouped in two large spikes around either end of the probability distribution, leaving only a few values between them. To unambiguously separate the galaxies that belong to each spike, we have applied a non-parametric binary classification scheme inspired on Otsu's method \citep{Otsu+1979} -- a one-dimensional discrete analogue of Fisher's Discriminant Analysis -- that in its simplest form splits the data into two classes by maximising the inter-class variance. Specifically, we have adopted the $95$ per cent of the maximum inter-class variance as the upper limit of this parameter that defines the ranges of values of $P_{\mathrm{PS}}$ that should be associated with each main spectral mode of the S0 galaxies, represented by the red and blue histograms.

This mapping in the PC1--PC2 plane which, as we have discussed above, is intimately connected to spectral parameters that inform on the activity of galaxies, bears some obvious similarities with the bimodal distribution into red and blue sequences shown by the general galaxy population in the colour-magnitude diagrams of optical broad-band photometric studies \citep[see e.g.][and references therein]{BM09}. This has prompted us to designate the thin, compact band delineated in the PC1--PC2 subspace by the S0 which essentially lack $\text{H}\,\alpha$ emission (shown in red colour too in the right panel of Fig.~\ref{fig:split}) as the \emph{Passive Sequence} (hence the PS subscript), whilst the name \emph{Active Cloud} (AC) is used to denote the much more disperse area encompassed by the objects with star formation signatures (identified with blue colour in Fig.~\ref{fig:split}). Likewise, the narrow strip of the PC1--PC2 subspace, which houses the few S0 that fall in the deep valley separating the two main modes of $P_{\mathrm{PS}}$ because of their in-between spectral characteristics (represented in gray in Fig.~\ref{fig:split}), has been named the \emph{Transition Region} (TR).

\begin{figure*}
	\includegraphics[width=\textwidth]{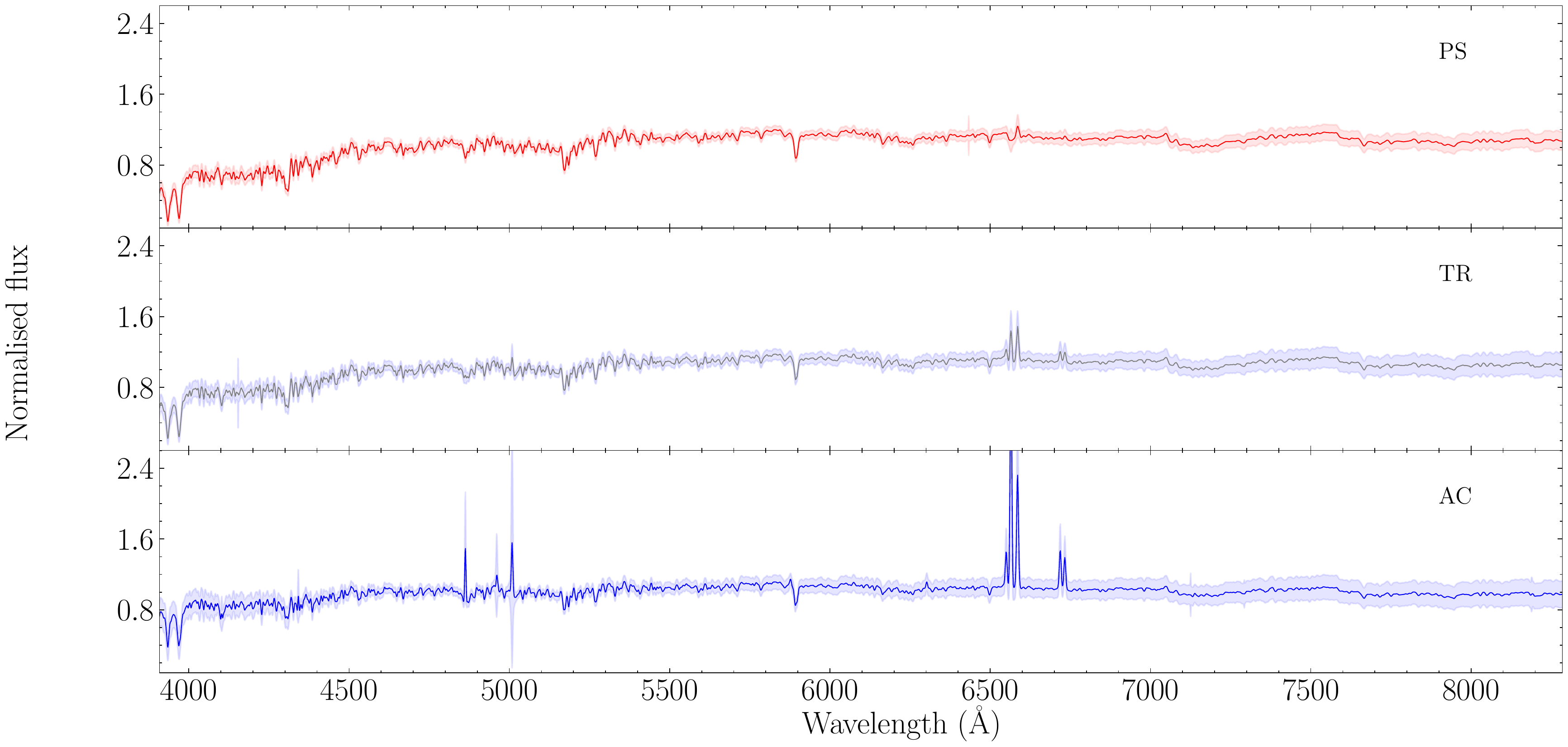}
    \caption{Mean spectra of the spectral classes of S0 determined from the first two principal components inferred in the PCA of our VLSS0 dataset. From top to bottom: the Passive Sequence (PS) in red colour, the Transition Region (TR) in grey, and the Active Cloud (AC) in blue. The shaded areas around the solid curves represent the $1\sigma$ variations about the average.}
    \label{fig:mean_f}
\end{figure*}
 
The suitability of the naming adopted to identify the different spectral classes detected in the PC1--PC2 subspace becomes evident when examining the mean spectrum of each group. In Fig.~\ref{fig:mean_f} we show, from top to bottom, the mean optical spectral flux distributions of the galaxies classified as PS (red spectrum), TR (grey) and AC (blue). The lack of emission lines, the deep $D4000$ \AA\ break and the rather red continuum seen in the mean spectrum of the PS are well-known characteristics of absorption-dominated galaxies, whereas the relatively strong emission lines, the shallower $D4000$ \AA\ break and the somewhat bluer continuum are traits that warrant our classification of the AC galaxies as objects with a significant level of activity, which seems to be mainly related to star formation (see next section). Note that the average spectrum of the latter class does not fit well within the classic picture of S0 galaxies, which are better represented by a spectrum similar to that of a typical PS object.

In a similar vein, we confirm the expectation that the characteristics of the mean spectrum of TR objects are intermediate with respect to those of the two main classes, with a relatively significant $\text{H}\,\alpha$ emission line, but with other emission lines from metals absent or rather weak. Fig.~\ref{fig:mean_f} also shows that the TR and AC spectral classes have $1\sigma$ uncertainties about the average (represented by the shaded areas surrounding the mean fluxes) bigger than that of the PS class, which reveal the greater variability shown by their spectra. 

\subsection{Physical properties of the spectral classes}
\label{S:phy_prop}

The total numbers and corresponding fractions of objects of the VLSS0 that make up the two main S0 sub-populations, PS and AC, identified by our classification scheme are listed in the second and third rows of Table~\ref{tab:summary}. The values in parentheses provide the same information for the whole MLSS0 dataset. It follows that about two thirds of the brightest S0 in the local universe are members of the Passive Sequence and show spectral characteristics which would agree with the classic image that is expected for these early-type galaxies. Nevertheless, we also have found that a quarter of the local S0 population belongs to the AC class\footnote{Given that the galaxies with emission lines appear to have bigger difficulties than quiescent objects in passing the filtering applied to select high-quality spectra (see Section~\ref{S:pca}), we cannot exclude the possibility that we are underestimating the true AC fraction somewhat.} and hence presents a richer spectra with relatively strong emission lines that demonstrates that not all S0 are necessarily quiescent objects.

Table~\ref{tab:summary} also summarises for each main spectral class the median values of the probability distribution functions (PDF) and their associated uncertainties, given in terms of interquartile ranges, of the eight properties most tightly correlated with the first principal components. The properties listed are (see also Section~\ref{S:proxy}): the equivalent width of the $\text{H}\,\alpha$ line, (row\ 5 in Table~\ref{tab:summary}), specific SFR (row\ 6), global SFR (row\ 7), stellar metallicity, represented by $Z$ (row\ 8), total stellar mass (row\ 9), stellar mass-to-$r$-band-light ratio (row\ 10), $(g - r)$ colour index (row\ 11), and $D4000$ \AA\ break (row\ 12). A more detailed inter-class comparison of these parameters is provided in Appendix~\ref{app:distributions}, where we display their full PDF in the form of violin plots.

It becomes clear from both the summary statistics reported in Table~\ref{tab:summary}, and the symmetric density plots of Fig.~\ref{fig:violins} that the two sub-populations of lenticulars show many important differences on their typical physical parameters. This is confirmed by the application of a two-sample Kolmogorov-Smirnov (KS) test to the eight pairs of PDF, which has produced p-values $\ll 0.05$ in all cases, indicating that there is a very small likelihood that the two types of S0 come from the same parent population. 

\begin{table*}
\centering
\caption{Amount of galaxies of the VLSS0 within each main spectral class, as well as medians and lower and upper quartiles of several of their most important physical properties.}
\label{tab:summary}
\begin{tabular}{|l|rrr|rrr|} 
\hline
S0 sub-population& \multicolumn{3}{c|}{Passive Sequence} & \multicolumn{3}{c|}{Active Cloud}  \\ 
\hline
$N$ (in MLSS0) & \multicolumn{3}{c|}{22067 (43579)}& \multicolumn{3}{c|}{7933 (20024)}  \\
$\%$ (in MLSS0) & \multicolumn{3}{c|}{69 (64)} & \multicolumn{3}{c|}{25 (29)} \\ 
\hline
Summary statistics  & $Q_1$ & Median  & $Q_4$ & $Q_1$ & Median  & $Q_4$ \\ 
\hline
$\log(\text{EW}(\text{H}\,\alpha)/$\AA) & -0.516   & -0.157  & 0.108 & 0.944 & 1.164   & 1.401 \\
$\log ({\text{sSFR}/yr^{-1}})$ & -12.317  & -11.934 & -11.409 & -10.600  & -10.327 & -10.033 \\
$\log ({\text{SFR}/M_{\sun}\ yr^{-1}})$ & -1.557 & -1.167  & -0.584 & 0.133 & 0.376 & 0.603 \\
$ Z/\text{Z}_{\sun}$ & 1.259 & 1.498 & 1.717 & 0.820 & 1.012 & 1.199 \\
$\log (M_\text{star}/{\text{M}_{\sun}})$ & 10.645 & 10.762  & 10.910 & 10.530 & 10.665  & 10.818 \\
${M_\text{star}/L_r}$ $({\text{M}_{\sun}/L_{\sun}})$ & 2.238 & 2.506 & 2.805 & 1.653 & 2.027 & 2.464 \\
$(g - r)$ & 0.713 & 0.740 & 0.766 & 0.587 & 0.641 & 0.687 \\
$D4000$ & 1.804 & 1.877 & 1.940 & 1.322 & 1.425 & 1.521 \\
\hline
\end{tabular}
\end{table*}

A further distinctive feature between the PDF of both spectral classes is that for PS objects the distributions of some of the properties investigated, especially the two SFR, $Z$, and to a lesser extent $\text{EW}(\text{H}\,\alpha)$, tend to be right-skewed and heavy-tailed (i.e.\ to have longer and heavier right tails), while the AC's distributions are more symmetric. Our examination of the robustness of the results rules out the possibility that this asymmetry obeys to a misclassification of highly-inclined AC members as PS galaxies. Therefore, this could point to the existence of a sub-class within the PS lenticulars that would have escaped detection.

\begin{figure*}
    \includegraphics[width=\textwidth]{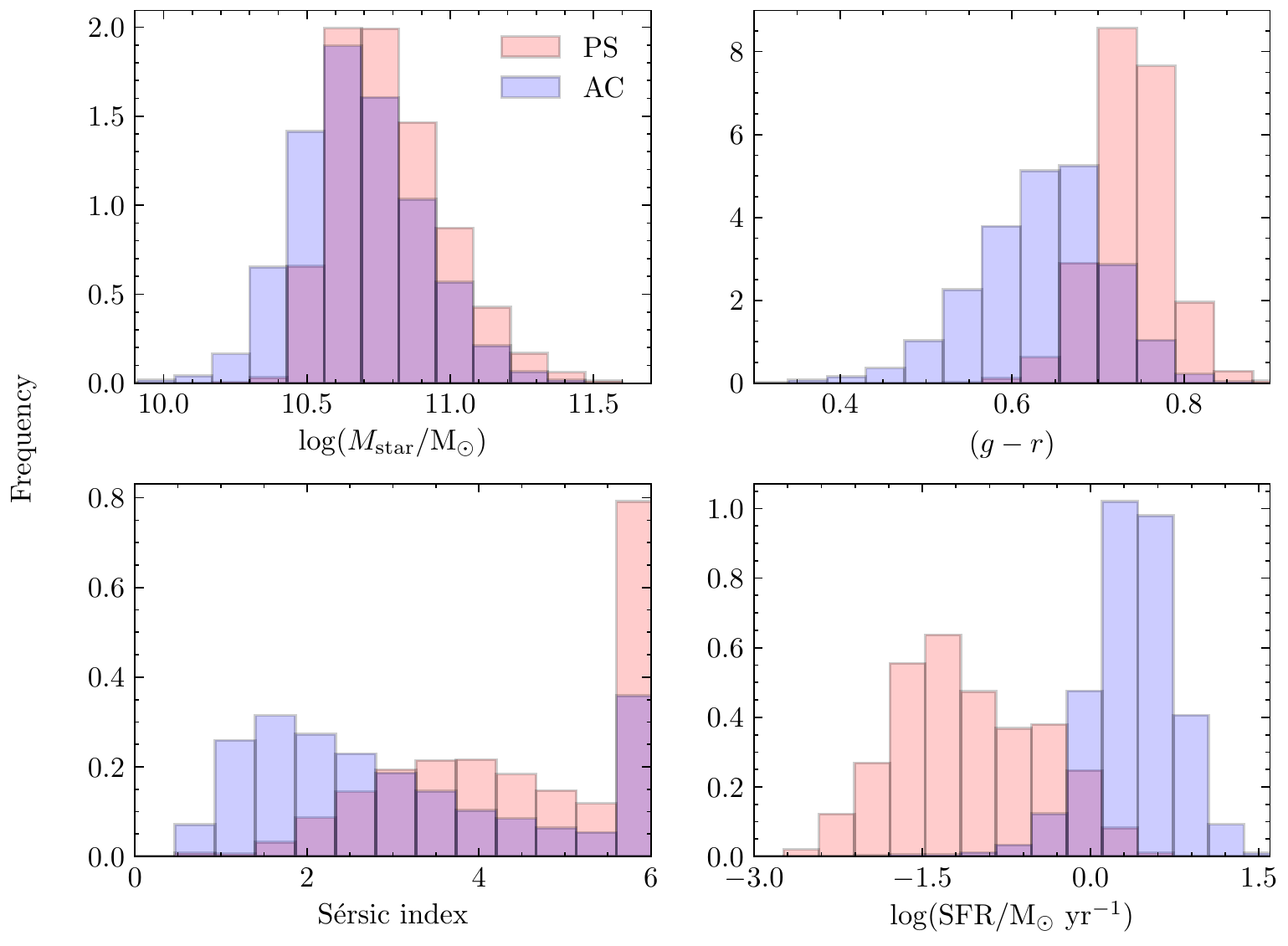}
    \caption{From left to right and top to bottom: histograms of the stellar mass, $(g-r)$ colour, global S\'ersic index, and star formation rate of lenticulars split by their spectral class. Red is used for Passive Sequence (PS) objects and blue for members of the Active Cloud (AC). According to a two-sample KS test, the two spectral classes show statistically significant differences on all these properties (the p-values are all $\ll 0.05$). The very high peaks of the S\'ersic index at the right end of the distributions are unphysical, as they arise from an artificial limitation in the calculation of this parameter to values lower than 6.}
\label{fig:4hist}    
\end{figure*}

\begin{figure}
    \includegraphics[width=\columnwidth]{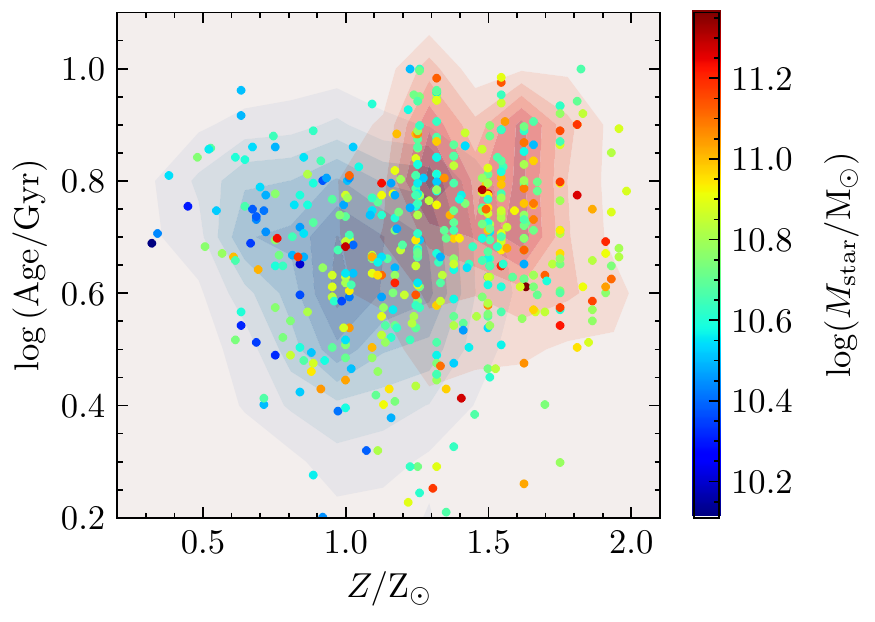}
    \caption{Contour plots showing the bivariate distribution of stellar ages and metallicities for galaxies in the VLSS0 partitioned into PS (red) and AC (blue) members, the darker the colour the higher the density. Three per cent of the data points in this sample, to avoid overcrowding, have also been included, colour-coded by their stellar mass. The graph shows that our spectral classification of the S0 is reasonably correlated with both the age and the metallicity of the stellar population of these galaxies. However, there is no evidence of a substantial correlation with stellar mass.}
\label{fig:age_met_mass}    
\end{figure}

To facilitate a more direct comparison between our results and those of previous studies of S0 galaxies that rank them according to some measurable global property, we plot in Fig.~\ref{fig:4hist} the histograms of the two S0 classes detected in our VLSS0 as a function of stellar mass (top-left), $(g-r)$ colour (top-right), global S\'ersic index (bottom-left), and SFR (bottom-right), respectively. Likewise, in Fig.~\ref{fig:age_met_mass} we provide for these same galaxies number density contours showing the bivariate distribution of their stellar ages and metallicities and plotting also some data points colour-coded by stellar mass. The histograms, in particular, allow one to better appreciate the different discriminant capabilities of various spectrophotometric measurements when assessing membership in one of the S0 sub-populations that we have identified. The most unambiguous division of the spectral classes would be provided by the SFR (or its per-unit-mass counterpart, sSFR), which is capable of separating the two main modes within much of the dynamic range of this property, proving almost as effective as our PCA-based technique. This is to be expected given the relatively direct bearing that our automated classification method has with those physical parameters more intimately related to star formation (indeed a very similar result would have also been obtained by using the $\text{EW}(\text{H}\,\alpha)$ or the $D4000$ break). For its part, the $(g-r)$ colour would perform well as divider on the extremes of its dynamic range, but could not resolve membership on its central part where the two S0 spectral modes show a substantial overlap. By contrast, the two sub-populations of S0 identified in our sample produce similar distributions of the global S\'ersic index and, especially, of the total stellar mass -- despite the long-recognised close relationship of the latter with the star formation history of galaxies \citep[e.g.][]{GS96} --, which would make these two properties unsuitable as discriminants of the two spectral modes. Recall, however, that in all the four cases the differences between the PS and AC distributions are highly significant according to the two-sample KS test.

As regards the characteristic values of the properties assigned to each spectral class, our results are largely consistent with the outcomes by \citet{Xiao+2016} and \citet{mckelvie+2018}, who using samples significantly smaller than ours find that star-forming S0 galaxies are less massive, bluer\footnote{\citet{mckelvie+2018} use Wide-Field Survey Explorer (WISE) mid-infrared colours as indicators of recent star formation activity.}, and possess shallower, i.e.\ less concentrated, central light profiles, with a possible prevalence of pseudo-bulge-like components since a large proportion of them are best fitted by S\'ersic profiles of index $n < 2$ \citetext{see also the results of an early study of E+S0 galaxies by \citealt{Helmboldt+2008}}. On top of that, our analysis reveals the existence of a good number of S0, \emph{both active and passive}, with measurable star formation rates. Among the latter, the observed values are certainly low and, when divided by the luminous mass of the parent galaxy, perfectly consistent with the sSFR values typical of the so-called 'non-star-forming red sequence of galaxies' \citep[see e.g.][and references therein]{KE12}. Thus, despite they all have ongoing star formation, its reduced intensity has not prevented these galaxies from being assigned the passive status. On the other hand, the majority of the S0 belonging to the AC show SFR that are comparable or higher than the current rate of the Milky Way and that translate into values of the sSFR that are just as strong as those found in actively star-forming late spirals of the same mass. These high levels of star formation move this class of lenticulars largely away from the traditional view of quiescent galaxies.

The only discrepancy with respect to the former studies is the relevance attributed to the stellar mass, which is seen as the main driver of the division of lenticulars into two sub-populations associated with the degree of star formation \citetext{see e.g.\ also \citealt{Barway+2013} who use as a divider the $K$-band luminosity} -- and by extension of the formation pathway taken\footnote{To us it makes more sense to think of S0 mass as the \emph{consequence} of the formation pathway followed and not as the cause.}, whilst, as noted above, it plays a lesser role in our investigation. Fig.~\ref{fig:age_met_mass}, which is similar to fig.\ 5 of \citet{mckelvie+2018}, supports further this conclusion, showing that, in contrast with those findings, in our dataset the stellar mass provides a division between the older, metal-rich PS objects and the younger, metal-poor AC counterparts, which is less clear. Differences in the mass ranges encompassed by the datasets analysed -- unlike the other studies, ours deals with a volume-limited sample that for this reason contains lenticulars of relatively high mass -- could explain the inconsistency.

By way of a summary, one can conclude that the S0 belonging to the AC sub-population differ from their PS counterparts in that the former:
\begin{enumerate}
\item are only slightly less massive on average, although somewhat more luminous, so they have lower $M_\text{star}/L$ ratios;
\item have a younger, bluer stellar population, that is ostensibly poorer in metals;
\item have lower global S\'ersic indices that follow a more asymmetric, positively skewed distribution of modal value $n\lesssim 2$; 
\item are almost entirely actively star-forming systems with average star formation rates, of about $2\;\text{M}_{\sun}\,\text{yr}^{-1}$, that are more than one order of magnitude higher.
\end{enumerate}

\subsection{Relationship with the environment}
\label{S:local_env}

\begin{figure}
    \includegraphics[width=\columnwidth]{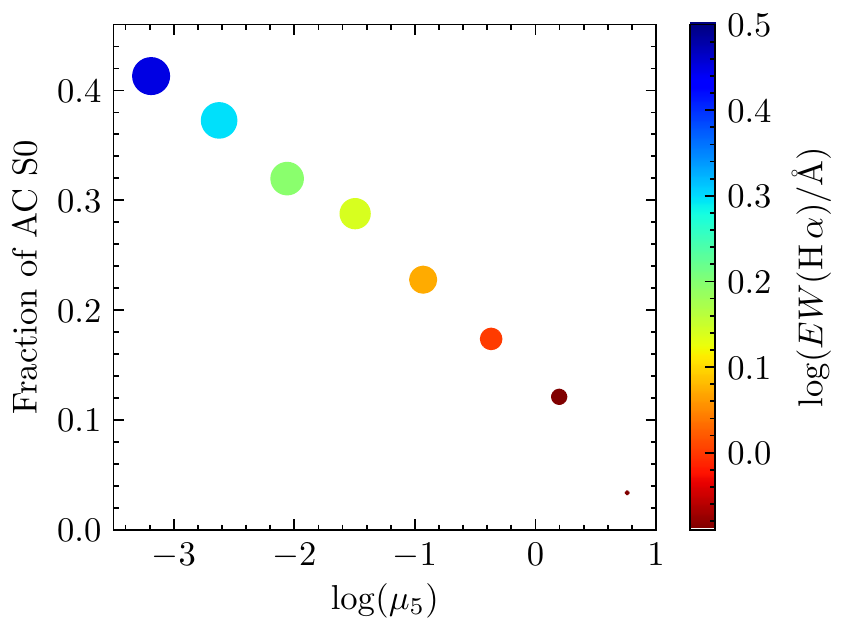}
    \caption{Number fraction of AC lenticulars as a function of the logarithm of the local density for all the galaxies in our VLSS0. The colour of the data points indicates the average value of $\log(\text{EW}(\text{H}\,\alpha))$ for the whole sample, whilst their size is proportional to the mean $\log(\text{SFR})$. This plot illustrates the dramatic impact the environment has on the activity of S0 galaxies.}
    \label{fig:density_frac}
\end{figure}

One key factor when it comes to understanding galaxy properties is the environment in which these objects reside. We have chosen to characterise it by means of the Galactic-extinction-corrected 3D galaxy number density estimator, $\mu_5$, defined in Section~\ref{S:localmu}. In Fig.~\ref{fig:density_frac}, we show the evolution in the number fraction of AC lenticulars with local density for all our VLSS0 galaxies, with the colour and size of the data points informing, respectively, about the corresponding change in the average equivalent width of the $\text{H}\,\alpha$ line and in the average SFR of the whole population. This plot reveals that while PS lenticulars are present in all kind of environments, the abundance of AC objects decreases linearly with increasing $\log(\mu_5)$, going from fractions above forty per cent at the lowest densities probed by our sample to being virtually absent in the densest regions of the local volume. (As in the case of their intrinsic physical properties, we find that the differences in the PDF of the local density for the two main spectral classes of lenticulars are statistically very significant, with a two-sample KS test returning $p \ll 0.05$.) Accordingly, the average values of the $\text{EW}(\text{H}\,\alpha)$ and SFR for the S0 population are also steadily reduced with local projected density (they follow a roughly linear decrement with $\log(\mu_5)$ as well), suggesting that the denser the environment the more severe is the quenching of the star-formation activity in S0 galaxies.  

Once again, our results show a good level of consensus with the outcomes of the studies mentioned in the previous section, as well as consistency with expectations for the general galaxy population \citep{Rines+2005}. This is the case of \citet{Xiao+2016}, who claim that active S0 are mainly located in the sparse environment, whilst quiescent lenticulars or with low-level star formation and/or AGN activity reside preferentially in the dense environment. \citet{Helmboldt+2008} also arrive at a similar conclusion with their samples of star-forming early-type and K+A galaxies, finding that both types of objects reach higher fractions among all galaxies in lower density environments. Only \citet{mckelvie+2018} seem to have an apparently discordant view when, from an analysis based on two environmental indicators -- one that measures the strength of the local tidal perturbations from neighbours and another that evaluates the large-scale structure in terms of the projected density to the 5th nearest neighbour --, find no significant evidences of a connection between the stellar masses, ages, and metallicities of MaNGA's lenticulars and their environment. These same authors, however, warn that their conclusion is based on an incomplete version of the MaNGA dataset in which the statistically rare cluster members are missing. Indeed, by contrast with the stellar mass, it can be observed that while our density measurements span over about four orders of magnitude and encompass the most extreme environments, the spatial spectroscopy data used in the former study, apart from being much fewer, sample a significantly narrower range of densities. Thus, it appears that the widely documented systematic differences observed in the (star-formation) activity between late-type spirals in rich environments and those in the field \citep[see e.g.][and references therein]{Bose16} are not exclusive to this end of the Hubble sequence, but they can actually be extended to all disc galaxies. This suggests that the physical properties of S0 galaxies are not detached from the environment in which they find themselves. 

\begin{figure*}
	\includegraphics[scale=1.1]{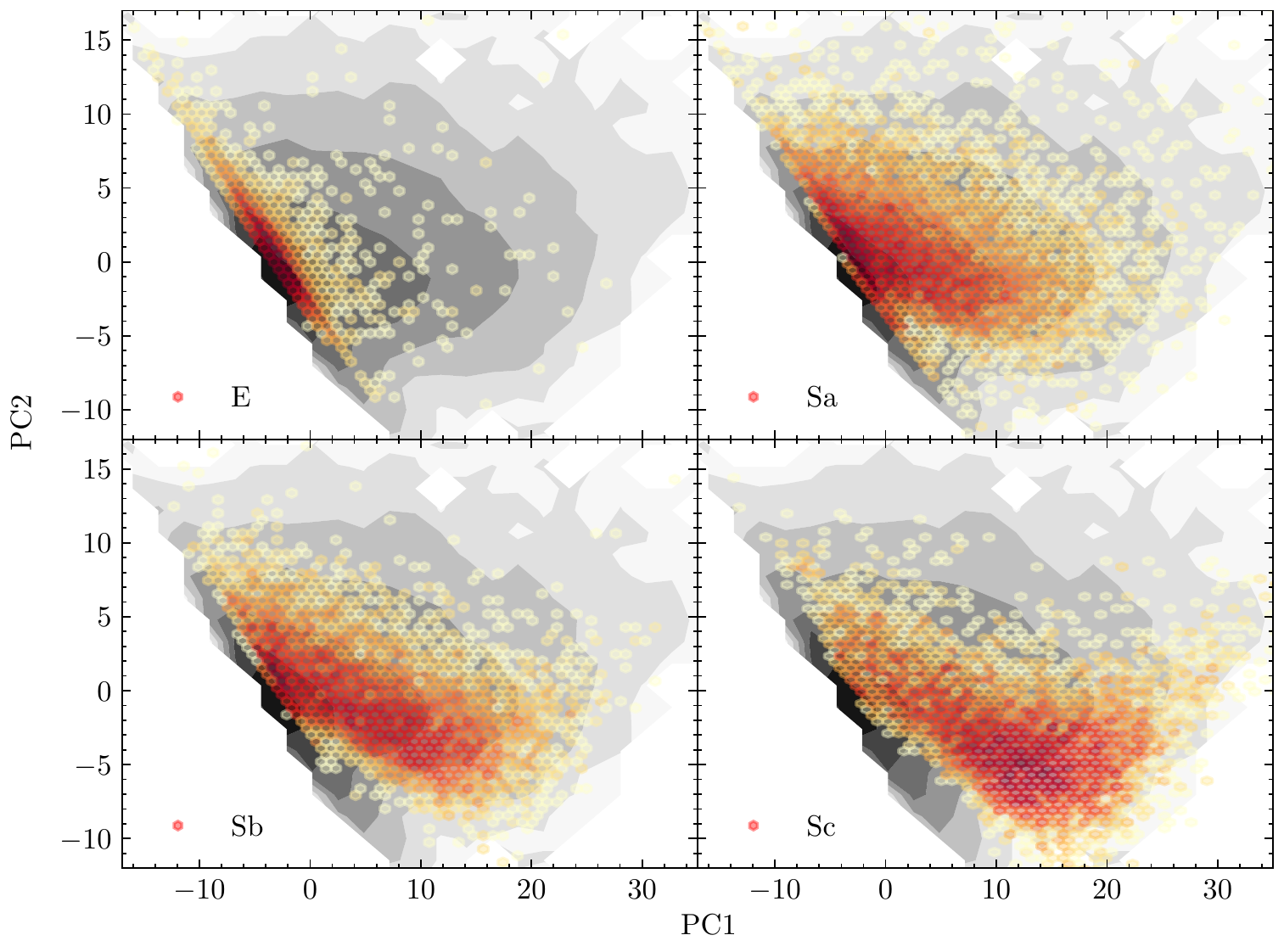}
    \caption{Number density distributions of galaxies of the main Hubble types in the PC1--PC2 plane defined by the S0. The background grey-scale contours represent equally spaced logarithmic densities for the lenticular population. Overlaid on top are the density distributions of E, Sa, Sb, and Sc galaxies calculated in this same subspace from arbitrary samples of $\sim 10,000$ spectra of each type \citetext{the morphologies are from the catalogue by \citealt{Dom&Sanch+2018}}. The colour intensity of the hexagonal cells is also graduated on a logarithmic scale, with the darker ones identifying the most populated bins. Note how the density distributions of bright galaxies change from being fundamentally in the PS to occupy mostly the bluest portion of the AC (see Fig.~\ref{fig:pc1_pc2_prop}) as the Hubble type goes from E to Sc. Note also that just as there are hardly any E distributed across the AC region, Sc galaxies are virtually absent from the PS.}
    \label{fig:s0_vs_oTT}
\end{figure*}

\section{Summary and discussion}
\label{S:summary}

In this work we have applied several Machine Learning techniques to analyse a sample of $68{,}043$ optical spectra of S0 galaxies in the local universe ($0.01 < z \lesssim 0.1$), by much the largest dataset of any kind ever assembled for this galaxy population. Our aim has been to investigate the physical properties of this type of objects in a more objective and informative way using their whole optical spectra, instead of the classical diagnostics based exclusively on photometric measurements or on a few specific spectral lines. Our analysis relies on both the spectra measured in the SDSS Legacy Survey and the morphological information listed in the recently published catalogue of automatically classified galaxies by \citet{Dom&Sanch+2018}. These data have been complemented with a variety of spectrophotometric parameters gathered from further external sources, including several parameters taken from the NSA and other well-known and much used SDSS-based catalogues. The final database has been completed with our own estimation of the local density corrected for the effects of Galactic interstellar extinction in a volume-limited subset of $32{,}188$ S0 galaxies with $M_r \lesssim -20.5$ mag.

We have used a Principal Component Analysis to extract the most relevant features of the S0 optical spectra. This objective and a priori-free technique has enabled us to reduce the highly multidimensional spectral data -- the number of pixels per spectrum is about 3800 for the SDSS-I/II spectra -- to a low-dimensional space that is optimal for this purpose. This is because it relies on the most important projections along a few new orthogonal axes, the eigenspectra (ES; also known as the Principal Components), that maximise the variance and, hence, minimise information loss. 

The fact that we can explain with only 3 ES nearly $95$ per cent of the variance shown by a training sample of S0 spectra is an indication that all the physics of the local population of S0 galaxies should be essentially contained in a few dimensions, probably not more than two if we stick to their passive representatives, due to the strong correlation shown by the first two principal components of this ensemble (see below). The first two ES, which already explain about 90 per cent of the variance, are characterised by the presence of strong emission lines, notably the $\text{H}\,\alpha$+[\ion{N}{II}]+[\ion{NS}{II}] complex, a few lines from highly ionised species such as [\ion{O}{III}], and higher-order Balmer lines up to $\text{H}\,\epsilon$. The most important differences lie in the continuum, blue for the first ES and red for the second, and in the presence of the \ion{Ca}{II} H+K doublet in emission in the latter. Many of these lines are also visible in the third ES, some of them now in absorption, which shows an essentially flat continuum. Preliminary analyses have revealed that this component, which only embodies six per cent of the spectral variance, is, however, a key diagnostic element for the identification of nuclear activity in S0 galaxies, an issue that we have left for future research.

The projections of the S0 spectra on the bivariate plane defined by the first two ES, denoted PC1 and PC2, have unveiled the existence of two main regions: a crowded narrow band with a sharply defined left edge that crosses the diagram diagonally, defining a zone in this plane in which the values of the first two predictors show a strong linear correlation, and a substantially less populated and much more scattered cloud of points that runs from the right of this band, where the PC scores are linearly uncorrelated. We have used a combination of a Logistic Regression and Otsu's method to perform an optimal separation of all galaxies in our dataset into two main classes according to their location in the PC1--PC2 subspace. The resulting spectral subsets have been respectively called the 'Passive Sequence' and the 'Active Cloud', as they include sub-populations of S0 with spectra representative of passive and active galaxies. In between these two distributions we have defined a narrow dividing zone that we have named the 'Transition Region', which contains objects with intermediate spectral characteristics. This classification of the S0 is markedly reminiscent of the 'Red Sequence/Green Valley/Blue Cloud' features of the well-known colour-magnitude diagram used to classify the entire galaxy population, thus the names adopted. Investigation of the potential biases that the fixed aperture of fibres and the inclination of galaxies (basically related to their internal extinction) could entail, has shown that the changes induced by these factors on the S0 spectra, and hence also on their PC scores, have, however, a negligible impact on their classification.

The differences in the spectra of each sub-population of S0 galaxies are also reflected in their different physical properties. Most of the intrinsic parameters analysed show a clear bimodality that correlates with the spectral division, giving weight to the notion that the two types of lenticulars that we have identified may have followed different evolutionary paths. This is especially evident for those attributes directly related to star formation, such as the SFR, sSFR, $\text{EW}(\text{H}\,\alpha)$, and $D4000$. However, it is remarkable that we do not see a stark separation in stellar mass, a property that some past studies \citep[e.g.][]{mckelvie+2018} have identified as the main driver of the division in lenticular sub-populations. We attribute this apparent inconsistency to the different mass ranges encompassed by the respective samples, since unlike these other studies, in the present work we have used a volume-limited subset of S0 galaxies that, for this reason, contains objects of relatively high mass. In any event, we have found that, for all the quantities investigated, the comparison between the PDF associated with each sub-population has always produced statistically significant differences. 

As regards the values of the properties assigned to each spectral class, our findings are, in general, in good agreement with the outcomes from most of the previous studies of this kind. Using a volume-limited subset of spectra, we have found that the members of the AC class tend to be slightly less massive than their passive counterparts, more luminous though with less concentrated light profiles, and to have a somewhat younger, bluer stellar component that is poorer in metals. Besides, while PS lenticulars are present in all kind of environments, AC objects inhabit preferentially sparser regions showing abundances that anti-correlate with the local density of galaxies. However, the most striking results are that virtually all AC lenticulars are actively star-forming systems, with average SFR comparable to those seen in late spiral galaxies -- in fact, they lie in the upper-mass range of the star-forming main sequence --, together with the observation that such systems may constitute, at least, a quarter of the S0 population in the local volume.

Indeed, by reducing the optical spectra of the other Hubble types to the most fundamental features inferred for the S0, one can see that the concordance between the AC lenticulars and late-type disc galaxies is not solely confined to their similar levels of star formation. Thus, as shown in Fig.~\ref{fig:s0_vs_oTT}, the projections of the spectra of the E galaxies in the same PC1--PC2 plane defined by the S0, occupy essentially the narrow, lens-shaped region delineated by the PS spectral class, whilst the projections of the spectra of the different spiral classes progressively move towards higher values of PC1 and lower values of PC2, gradually shifting the peak of the corresponding density distributions from the origin of the plane to the lower-right corner of the AC's region. Even so, it can be seen that, in addition to all the spectra of E and Sa, virtually all of the spectra of Sb and the vast majority of those of Sc remain in this subspace within the limits delineated by the lenticular population. All this can be taken as evidence that the traditional conception of present-day S0 as basically red and dead galaxies is not correct, and that this morphological group should increasingly be treated as an heterogeneous collection of objects \citep[see also][]{WS03,Mor06,Cro10,Barway+2013}.

\begin{figure}
    \includegraphics[width=\columnwidth]{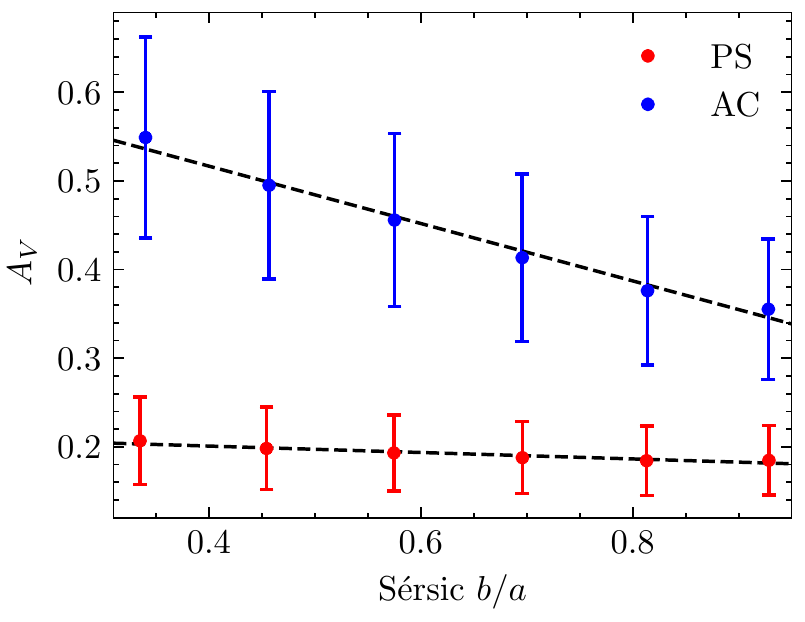}
    \caption{Inclination dependence of the total internal extinction in the photometric $V$ band for the two main spectral classes of S0. The data points and bars represent, respectively, the means and $1\sigma$ errors of $A_V$. The axial ratio $b/a$ derived by the SDSS from two-dimensional, single-component S\'ersic fits in $r$-band is used as a proxy for inclination. The adopted range of values for this observational parameter takes into account the results by \citet{Masters+10}, who find that the intrinsic $b/a$ of S0 galaxies is 0.23.}
    \label{fig:Avvsba}
\end{figure}

Implicit in the fact that one of the major distinguishing attributes of the two main spectral classes of lenticulars is the level at which they are currently forming stars, is the idea that one should expect important differences in their ISM too. A simple way of checking this would be to use the measurements of the HI content from the ALFALFA survey included in our database. In practice, however, this is not feasible as the fraction of S0 that have 21-cm data available is very small (our dataset contains just a few hundred HI observations for S0), mostly due to the substantially lower depth of ALFALFA ($z_{\text{hel}} \lesssim 0.06$) and its only partial overlap with the SDSS footprint. Instead, we use another effective way of revealing the presence of diffuse matter in galaxies, such as their internal extinction of starlight, for which we do have abundant data. Fig.~\ref{fig:Avvsba} demonstrates that our assessment on the ISM of lenticulars is correct. In this plot we compare the inclination dependence of the total internal extinction in the $V$ (550 nm) band, $A_V$, for the two main S0 sub-populations. The differences are striking: while the PS galaxies show a mild, and essentially inclination-free, nebular extinction, the degree of extinction of AC members is significantly higher and grows linearly with inclination. The fact that we obtain such markedly different outcomes for the two types of S0 using a purely photometric parameter reinforces the reliability of our spectral classification.

The study of the ISM of S0 has been completed with the application of the statistical test by \citet{JDT96} geared towards the determination of the typical optical depth of a sample. This qualitative test consists of: i) inferring the observed distribution of inclinations, using the ratio of the semi-minor to semi-major axes; ii) comparing it with different expectations from simulated samples of disc galaxies affected by the same selection effects and that contain objects that are either completely optically thin or completely optically thick; and iii) determining which theoretical distribution best corresponds to the actual data. We show in Fig.~\ref{fig:inclination_hist} the observed distributions of inclinations for the two spectral classes of lenticulars included in the whole MLSS0 dataset, which is a magnitude-selected sample. Comparison of these distributions with fig.~1b of these authors reveals that the S0 can be considered, in general, galaxies with a relatively low opacity, although the AC's members appear to be, as expected, optically thicker than their PS counterparts, which behave as essentially transparent discs. The possible effects of internal extinction, as well as of the fixed aperture of the fibres, on the spectra of the S0 have been explored in detail on Appendix~\ref{app:robustness}, where we demonstrate that they have a marginal impact on our PCA-based classification of these objects. 

\begin{figure}
     \includegraphics[width=\columnwidth]{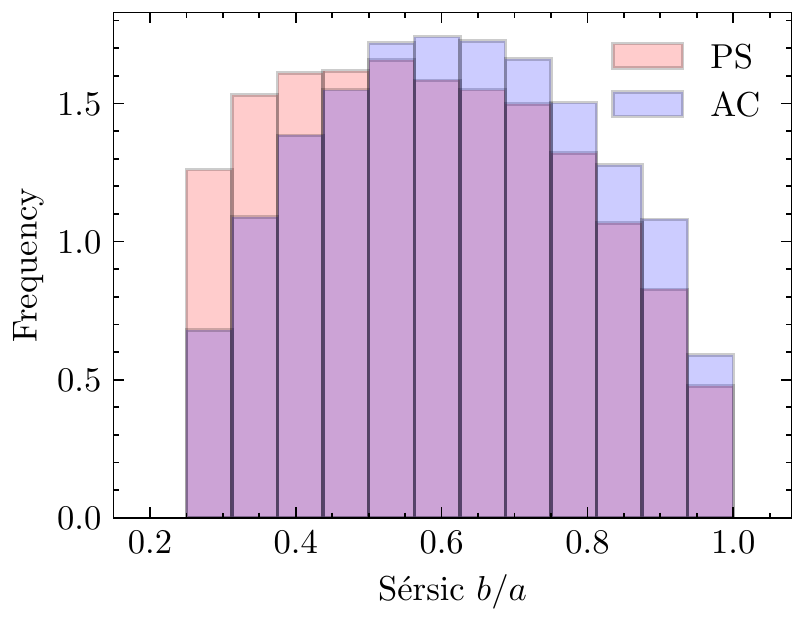}
     \caption{Observed distribution of inclinations for the two main spectral sub-populations of lenticulars in the MLSS0.}
     \label{fig:inclination_hist}
\end{figure}

At this point, we feel that it is still premature to start too detailed a discussion about the possible formation paths followed by the two sub-populations of lenticulars that we have uncovered. In relation to this issue, one of our most relevant results is the very significant segregation found between the intrinsic attributes of the two spectral classes of S0. The nature of the differences is such that prevents PS lenticulars to be the outcome of the internal secular evolution of AC objects, in the sense that the PDF of some properties, such as $M_\text{star}$, are at odds with a closed-box transformation of the local AC into the quiescent S0. However, if one takes into account the downsizing paradigm in its appearance of the decline with time of the typical mass of star-forming galaxies \citep[e.g.][]{Fontanot+2009}, then the PS members could well be the descendants of intermediate-$z$ lenticulars with AC characteristics, which could even have also acted as progenitors of the spheroid-dominated post-starburst K+A galaxies that reside preferentially in the field. Implicit in this scenario is an increase of the AC/PS fraction with redshift. On the other hand, the clear anti-correlation found between the local density of the environment and the fractional abundance of actively star-forming S0, which obeys a nearly perfect power law, is another of the elements that could shed some light in our understanding of the physics involved in S0 formation. This relationship can be naturally explained in cluster regions where hydrodynamic interactions are expected to drive the connection of galaxy properties with the radial run of the density in these systems. However, it does not seem very likely that this sort of mechanisms are solely responsible for a correspondence that, as we have shown, extends from the densely populated regions of the local volume all the way down to the most sparse environment. There, we still see that the galaxies bearing the S0 designation are mostly absorption-dominated objects, which are unlikely to form from gas processes but rather through gravitational effects. It is even possible that these latter mechanisms do not even work in the emptiest parts of space, where the density of galaxies is too low for their mutual interactions to be really effective. 

Taken as a whole, what all these results do indeed indicate is that the environment of S0 galaxies must play a pivotal role in the modulation of their star formation activity and, possibly, also of many other of their physical properties. But with the information we have gathered to date, we cannot answer yet the fundamental question of what are the specific environmental mechanisms involved and how have them played out in detail, without going too far into the territory of speculation. However, our study of the S0 populations is still in its infancy. Indeed, we will continue working in this area in the hope that our research may reveal new evidence in a not too distant future that allows us to draw well-founded and more specific conclusions about the possible formation scenarios of these fascinating objects.

\section*{Acknowledgements}

The authors thank the reviewer for his/her detailed, insightful and encouraging comments, which have enabled us to refine the focus of the paper. We acknowledge financial support from the Spanish AEI and European FEDER funds through the research project AYA2016--76682--C3. J.L.T.\ was also partially supported by a collaboration scholarship from the Royal Academy of Sciences and Arts of Barcelona and the University of Barcelona. Additional funding for this work has been provided by the State Agency for Research of the Spanish MCIU through the Centre of Excellence Severo Ochoa's award for the Instituto de Astrof\'\i sica de Andaluc\'\i a under contract SEV--2017--0709. This research has made use of data from the following databases: NASA-Sloan Atlas at \url{http://nsatlas.org/}, Morphological catalogue for SDSS galaxies by \citet{Dom&Sanch+2018}, GALEX-SDSS-WISE at \url{http://pages.iu.edu/~salims/gswlc/}, and both Portsmouth Stellar Kinematics and Emission Line Fluxes and eBOSS Firefly at \url{https://www.sdss.org/dr15/data_access/value-added-catalogs/}. We are grateful to all the people involved in the gathering, reduction, and processing of the data contained in all these catalogues, as well as the public and private institutions that have provided the necessary funding, resources and technical support to make possible both all these surveys and the release of their findings to the community.


\bibliographystyle{mnras}
\bibliography{biblio}

\begin{thebibliography}{}
\makeatletter
\relax
\def\mn@urlcharsother{\let\do\@makeother \do\$\do\&\do\#\do\^\do\_\do\%\do\~}
\def\mn@doi{\begingroup\mn@urlcharsother \@ifnextchar [ {\mn@doi@}
  {\mn@doi@[]}}
\def\mn@doi@[#1]#2{\def\@tempa{#1}\ifx\@tempa\@empty \href
  {http://dx.doi.org/#2} {doi:#2}\else \href {http://dx.doi.org/#2} {#1}\fi
  \endgroup}
\def\mn@eprint#1#2{\mn@eprint@#1:#2::\@nil}
\def\mn@eprint@arXiv#1{\href {http://arxiv.org/abs/#1} {{\tt arXiv:#1}}}
\def\mn@eprint@dblp#1{\href {http://dblp.uni-trier.de/rec/bibtex/#1.xml}
  {dblp:#1}}
\def\mn@eprint@#1:#2:#3:#4\@nil{\def\@tempa {#1}\def\@tempb {#2}\def\@tempc
  {#3}\ifx \@tempc \@empty \let \@tempc \@tempb \let \@tempb \@tempa \fi \ifx
  \@tempb \@empty \def\@tempb {arXiv}\fi \@ifundefined
  {mn@eprint@\@tempb}{\@tempb:\@tempc}{\expandafter \expandafter \csname
  mn@eprint@\@tempb\endcsname \expandafter{\@tempc}}}

\bibitem[\protect\citeauthoryear{{Alam} et~al.,}{{Alam}
  et~al.}{2015}]{Alam+2015}
{Alam} S.,  et~al., 2015, \mn@doi [\apjs] {10.1088/0067-0049/219/1/12}, \href
  {https://ui.adsabs.harvard.edu/abs/2015ApJS..219...12A} {219, 12}

\bibitem[\protect\citeauthoryear{{Allen} et~al.,}{{Allen} et~al.}{2015}]{All15}
{Allen} J.~T.,  et~al., 2015, \mn@doi [\mnras] {10.1093/mnras/stu2057}, \href
  {https://ui.adsabs.harvard.edu/abs/2015MNRAS.446.1567A} {446, 1567}

\bibitem[\protect\citeauthoryear{{Baldwin}, {Phillips}  \&
  {Terlevich}}{{Baldwin} et~al.}{1981}]{Baldwin+1981}
{Baldwin} J.~A.,  {Phillips} M.~M.,   {Terlevich} R.,  1981, \mn@doi [\pasp]
  {10.1086/130766}, \href
  {https://ui.adsabs.harvard.edu/abs/1981PASP...93....5B} {93, 5}

\bibitem[\protect\citeauthoryear{{Barnes}}{{Barnes}}{1999}]{Barnes+1999}
{Barnes} J.~E.,  1999, in {Barnes} J.~E.,  {Sanders} D.~B.,  eds,  IAU
  Symposium Vol. 186, Galaxy Interactions at Low and High Redshift. p.~137
  (\mn@eprint {} {astro-ph/9811091})

\bibitem[\protect\citeauthoryear{{Barway}, {Kembhavi}, {Wadadekar}, {Ravikumar}
   \& {Mayya}}{{Barway} et~al.}{2007}]{Barway+2007}
{Barway} S.,  {Kembhavi} A.,  {Wadadekar} Y.,  {Ravikumar} C.~D.,   {Mayya}
  Y.~D.,  2007, \mn@doi [\apj] {10.1086/518422}, \href
  {https://ui.adsabs.harvard.edu/abs/2007ApJ...661L..37B} {661, L37}

\bibitem[\protect\citeauthoryear{{Barway}, {Wadadekar}, {Vaghmare}  \&
  {Kembhavi}}{{Barway} et~al.}{2013}]{Barway+2013}
{Barway} S.,  {Wadadekar} Y.,  {Vaghmare} K.,   {Kembhavi} A.~K.,  2013,
  \mn@doi [\mnras] {10.1093/mnras/stt478}, \href
  {https://ui.adsabs.harvard.edu/abs/2013MNRAS.432..430B} {432, 430}

\bibitem[\protect\citeauthoryear{{Bedregal}, {Arag{\'o}n-Salamanca}  \&
  {Merrifield}}{{Bedregal} et~al.}{2006}]{Bedregal+2006}
{Bedregal} A.~G.,  {Arag{\'o}n-Salamanca} A.,   {Merrifield} M.~R.,  2006,
  \mn@doi [\mnras] {10.1111/j.1365-2966.2006.11031.x}, \href
  {https://ui.adsabs.harvard.edu/abs/2006MNRAS.373.1125B} {373, 1125}

\bibitem[\protect\citeauthoryear{{Blanton} \& {Moustakas}}{{Blanton} \&
  {Moustakas}}{2009}]{BM09}
{Blanton} M.~R.,  {Moustakas} J.,  2009, \mn@doi [\araa]
  {10.1146/annurev-astro-082708-101734}, \href
  {https://ui.adsabs.harvard.edu/abs/2009ARA&A..47..159B} {47, 159}

\bibitem[\protect\citeauthoryear{{Blanton}, {Kazin}, {Muna}, {Weaver}  \&
  {Price-Whelan}}{{Blanton} et~al.}{2011}]{Blanton+2011}
{Blanton} M.~R.,  {Kazin} E.,  {Muna} D.,  {Weaver} B.~A.,   {Price-Whelan} A.,
   2011, \mn@doi [\aj] {10.1088/0004-6256/142/1/31}, \href
  {https://ui.adsabs.harvard.edu/abs/2011AJ....142...31B} {142, 31}

\bibitem[\protect\citeauthoryear{{Blanton} et~al.,}{{Blanton}
  et~al.}{2017}]{Blanton+2017}
{Blanton} M.~R.,  et~al., 2017, \mn@doi [\aj] {10.3847/1538-3881/aa7567}, \href
  {https://ui.adsabs.harvard.edu/abs/2017AJ....154...28B} {154, 28}

\bibitem[\protect\citeauthoryear{{Boselli} et~al.,}{{Boselli}
  et~al.}{2011}]{Boselli+2011}
{Boselli} A.,  et~al., 2011, \mn@doi [\aap] {10.1051/0004-6361/201016389},
  \href {https://ui.adsabs.harvard.edu/abs/2011A&A...528A.107B} {528, A107}

\bibitem[\protect\citeauthoryear{{Boselli} et~al.,}{{Boselli}
  et~al.}{2016}]{Bose16}
{Boselli} A.,  et~al., 2016, \mn@doi [\aap] {10.1051/0004-6361/201629221},
  \href {https://ui.adsabs.harvard.edu/abs/2016A&A...596A..11B} {596, A11}

\bibitem[\protect\citeauthoryear{{Bundy} et~al.,}{{Bundy}
  et~al.}{2015}]{Bundy+2015}
{Bundy} K.,  et~al., 2015, \mn@doi [\apj] {10.1088/0004-637X/798/1/7}, \href
  {https://ui.adsabs.harvard.edu/abs/2015ApJ...798....7B} {798, 7}

\bibitem[\protect\citeauthoryear{Burstein, Ho, Huchra  \& Macri}{Burstein
  et~al.}{2005}]{Burstein+2005}
Burstein D.,  Ho L.~C.,  Huchra J.~P.,   Macri L.~M.,  2005, The Astrophysical
  Journal, 621, 246

\bibitem[\protect\citeauthoryear{{Butcher} \& {Oemler}}{{Butcher} \&
  {Oemler}}{1978}]{Butcher&Oemler+1978}
{Butcher} H.,  {Oemler} Jr. A.,  1978, \mn@doi [\apj] {10.1086/155751}, \href
  {http://adsabs.harvard.edu/abs/1978ApJ...219...18B} {219, 18}

\bibitem[\protect\citeauthoryear{{Cappellari} et~al.,}{{Cappellari}
  et~al.}{2011}]{Cap11}
{Cappellari} M.,  et~al., 2011, \mn@doi [Monthly Notices of the Royal
  Astronomical Society] {10.1111/j.1365-2966.2011.18600.x}, \href
  {https://ui.adsabs.harvard.edu/abs/2011MNRAS.416.1680C} {416, 1680}

\bibitem[\protect\citeauthoryear{{Casertano} \& {Hut}}{{Casertano} \&
  {Hut}}{1985}]{Casertano&Hut1985}
{Casertano} S.,  {Hut} P.,  1985, \mn@doi [\apj] {10.1086/163589}, \href
  {https://ui.adsabs.harvard.edu/abs/1985ApJ...298...80C} {298, 80}

\bibitem[\protect\citeauthoryear{{Colless} et~al.,}{{Colless}
  et~al.}{2001}]{Colless+2001}
{Colless} M.,  et~al., 2001, \mn@doi [\mnras]
  {10.1046/j.1365-8711.2001.04902.x}, \href
  {https://ui.adsabs.harvard.edu/abs/2001MNRAS.328.1039C} {328, 1039}

\bibitem[\protect\citeauthoryear{{Comparat} et~al.,}{{Comparat}
  et~al.}{2017}]{Comparat+2017}
{Comparat} J.,  et~al., 2017, arXiv e-prints, \href
  {https://ui.adsabs.harvard.edu/abs/2017arXiv171106575C} {p. arXiv:1711.06575}

\bibitem[\protect\citeauthoryear{{Connolly} \& {Szalay}}{{Connolly} \&
  {Szalay}}{1999}]{Connolly}
{Connolly} A.~J.,  {Szalay} A.~S.,  1999, \mn@doi [\aj] {10.1086/300839}, \href
  {https://ui.adsabs.harvard.edu/\#abs/1999AJ....117.2052C} {117, 2052}

\bibitem[\protect\citeauthoryear{{Couch}, {Ellis}, {Sharples}  \&
  {Smail}}{{Couch} et~al.}{1994}]{Cou94}
{Couch} W.~J.,  {Ellis} R.~S.,  {Sharples} R.~M.,   {Smail} I.,  1994, \mn@doi
  [\apj] {10.1086/174387}, \href
  {https://ui.adsabs.harvard.edu/abs/1994ApJ...430..121C} {430, 121}

\bibitem[\protect\citeauthoryear{{Couch}, {Barger}, {Smail}, {Ellis}  \&
  {Sharples}}{{Couch} et~al.}{1998}]{Couch+1998}
{Couch} W.~J.,  {Barger} A.~J.,  {Smail} I.,  {Ellis} R.~S.,   {Sharples}
  R.~M.,  1998, \mn@doi [\apj] {10.1086/305462}, \href
  {https://ui.adsabs.harvard.edu/abs/1998ApJ...497..188C} {497, 188}

\bibitem[\protect\citeauthoryear{{Crocker}, {Bureau}, {Young}  \&
  {Combes}}{{Crocker} et~al.}{2010}]{Cro10}
{Crocker} A.~F.,  {Bureau} M.,  {Young} L.~M.,   {Combes} F.,  2010, \mn@doi
  [\mnras] {10.1111/j.1365-2966.2010.17537.x}, 410, 1197

\bibitem[\protect\citeauthoryear{{Davis}, {Greene}, {Ma}, {Pand ya},
  {Blakeslee}, {McConnell}  \& {Thomas}}{{Davis} et~al.}{2016}]{Davis+2016}
{Davis} T.~A.,  {Greene} J.,  {Ma} C.-P.,  {Pand ya} V.,  {Blakeslee} J.~P.,
  {McConnell} N.,   {Thomas} J.,  2016, \mn@doi [\mnras]
  {10.1093/mnras/stv2313}, \href
  {https://ui.adsabs.harvard.edu/abs/2016MNRAS.455..214D} {455, 214}

\bibitem[\protect\citeauthoryear{{Dobos}, {Csabai}, {Yip}, {Budav{\'a}ri},
  {Wild}  \& {Szalay}}{{Dobos} et~al.}{2012}]{Dobos}
{Dobos} L.,  {Csabai} I.,  {Yip} C.-W.,  {Budav{\'a}ri} T.,  {Wild} V.,
  {Szalay} A.~S.,  2012, \mn@doi [\mnras] {10.1111/j.1365-2966.2011.20109.x},
  \href {http://adsabs.harvard.edu/abs/2012MNRAS.420.1217D} {420, 1217}

\bibitem[\protect\citeauthoryear{{Dom{\'\i}nguez S{\'a}nchez},
  {Huertas-Company}, {Bernardi}, {Tuccillo}  \& {Fischer}}{{Dom{\'\i}nguez
  S{\'a}nchez} et~al.}{2018}]{Dom&Sanch+2018}
{Dom{\'\i}nguez S{\'a}nchez} H.,  {Huertas-Company} M.,  {Bernardi} M.,
  {Tuccillo} D.,   {Fischer} J.~L.,  2018, \mn@doi [\mnras]
  {10.1093/mnras/sty338}, \href
  {https://ui.adsabs.harvard.edu/\#abs/2018MNRAS.476.3661D} {476, 3661}

\bibitem[\protect\citeauthoryear{{Dressler}}{{Dressler}}{1980}]{Dressler1980}
{Dressler} A.,  1980, \mn@doi [\apj] {10.1086/157753}, \href
  {https://ui.adsabs.harvard.edu/abs/1980ApJ...236..351D} {236, 351}

\bibitem[\protect\citeauthoryear{{Dressler} et~al.,}{{Dressler}
  et~al.}{1997}]{Dressler+1997}
{Dressler} A.,  et~al., 1997, \mn@doi [\apj] {10.1086/304890}, \href
  {https://ui.adsabs.harvard.edu/abs/1997ApJ...490..577D} {490, 577}

\bibitem[\protect\citeauthoryear{{Eliche-Moral}, {Rodr{\'\i}guez-P{\'e}rez},
  {Borlaff}, {Querejeta}  \& {Tapia}}{{Eliche-Moral}
  et~al.}{2018}]{Eliche-Moral+2018}
{Eliche-Moral} M.~C.,  {Rodr{\'\i}guez-P{\'e}rez} C.,  {Borlaff} A.,
  {Querejeta} M.,   {Tapia} T.,  2018, \mn@doi [\aap]
  {10.1051/0004-6361/201832911}, \href
  {https://ui.adsabs.harvard.edu/abs/2018A&A...617A.113E} {617, A113}

\bibitem[\protect\citeauthoryear{{Fasano}, {Poggianti}, {Couch}, {Bettoni},
  {Kj{\ae}rgaard}  \& {Moles}}{{Fasano} et~al.}{2000}]{Fasano+2000}
{Fasano} G.,  {Poggianti} B.~M.,  {Couch} W.~J.,  {Bettoni} D.,
  {Kj{\ae}rgaard} P.,   {Moles} M.,  2000, \mn@doi [\apj] {10.1086/317047},
  \href {https://ui.adsabs.harvard.edu/abs/2000ApJ...542..673F} {542, 673}

\bibitem[\protect\citeauthoryear{{Fischer}, {Dom{\'\i}nguez S{\'a}nchez}  \&
  {Bernardi}}{{Fischer} et~al.}{2019}]{Fischer+2019}
{Fischer} J.~L.,  {Dom{\'\i}nguez S{\'a}nchez} H.,   {Bernardi} M.,  2019,
  \mn@doi [\mnras] {10.1093/mnras/sty3135}, \href
  {https://ui.adsabs.harvard.edu/abs/2019MNRAS.483.2057F} {483, 2057}

\bibitem[\protect\citeauthoryear{{Fitzpatrick}}{{Fitzpatrick}}{1999}]{Fitzpatrick1999}
{Fitzpatrick} E.~L.,  1999, \mn@doi [\pasp] {10.1086/316293}, \href
  {https://ui.adsabs.harvard.edu/abs/1999PASP..111...63F} {111, 63}

\bibitem[\protect\citeauthoryear{{Fontanot}, {De Lucia}, {Monaco}, {Somerville}
   \& {Santini}}{{Fontanot} et~al.}{2009}]{Fontanot+2009}
{Fontanot} F.,  {De Lucia} G.,  {Monaco} P.,  {Somerville} R.~S.,   {Santini}
  P.,  2009, \mn@doi [\mnras] {10.1111/j.1365-2966.2009.15058.x}, \href
  {https://ui.adsabs.harvard.edu/abs/2009MNRAS.397.1776F} {397, 1776}

\bibitem[\protect\citeauthoryear{{Fraser-McKelvie}, {Arag{\'o}n-Salamanca},
  {Merrifield}, {Tabor}, {Bernardi}, {Drory}, {Parikh}  \&
  {Argudo-Fern{\'a}ndez}}{{Fraser-McKelvie} et~al.}{2018}]{mckelvie+2018}
{Fraser-McKelvie} A.,  {Arag{\'o}n-Salamanca} A.,  {Merrifield} M.,  {Tabor}
  M.,  {Bernardi} M.,  {Drory} N.,  {Parikh} T.,   {Argudo-Fern{\'a}ndez} M.,
  2018, \mn@doi [\mnras] {10.1093/mnras/sty2563}, \href
  {https://ui.adsabs.harvard.edu/\#abs/2018MNRAS.481.5580F} {481, 5580}

\bibitem[\protect\citeauthoryear{{Gavazzi} \& {Scodeggio}}{{Gavazzi} \&
  {Scodeggio}}{1996}]{GS96}
{Gavazzi} G.,  {Scodeggio} M.,  1996, \aap, \href
  {https://ui.adsabs.harvard.edu/abs/1996A&A...312L..29G} {312, L29}

\bibitem[\protect\citeauthoryear{{Giovanelli}, {Haynes}  \&
  {Chincarini}}{{Giovanelli} et~al.}{1986}]{GHC86}
{Giovanelli} R.,  {Haynes} M.~P.,   {Chincarini} G.~L.,  1986, \mn@doi [The
  Astrophysical Journal] {10.1086/163784}, \href
  {https://ui.adsabs.harvard.edu/abs/1986ApJ...300...77G} {300, 77}

\bibitem[\protect\citeauthoryear{{Giovanelli} et~al.,}{{Giovanelli}
  et~al.}{2005}]{Giovanelli+2005}
{Giovanelli} R.,  et~al., 2005, \mn@doi [\aj] {10.1086/497431}, \href
  {https://ui.adsabs.harvard.edu/abs/2005AJ....130.2598G} {130, 2598}

\bibitem[\protect\citeauthoryear{{Goto}, {Yamauchi}, {Fujita}, {Okamura},
  {Sekiguchi}, {Smail}, {Bernardi}  \& {Gomez}}{{Goto} et~al.}{2003}]{Got03}
{Goto} T.,  {Yamauchi} C.,  {Fujita} Y.,  {Okamura} S.,  {Sekiguchi} M.,
  {Smail} I.,  {Bernardi} M.,   {Gomez} P.~L.,  2003, \mn@doi [Monthly Notices
  of the Royal Astronomical Society] {10.1046/j.1365-2966.2003.07114.x}, \href
  {https://ui.adsabs.harvard.edu/abs/2003MNRAS.346..601G} {346, 601}

\bibitem[\protect\citeauthoryear{{Graham}}{{Graham}}{2014}]{Graham2014}
{Graham} A.~W.,  2014, {Scaling Laws in Disk Galaxies}.
p.~185

\bibitem[\protect\citeauthoryear{{Gunn} \& {Gott}}{{Gunn} \&
  {Gott}}{1972}]{Gunn&Gott1972}
{Gunn} J.~E.,  {Gott} J.~Richard I.,  1972, \mn@doi [\apj] {10.1086/151605},
  \href {https://ui.adsabs.harvard.edu/abs/1972ApJ...176....1G} {176, 1}

\bibitem[\protect\citeauthoryear{{Haynes} et~al.,}{{Haynes}
  et~al.}{2018}]{Haynes+2018}
{Haynes} M.~P.,  et~al., 2018, \mn@doi [\apj] {10.3847/1538-4357/aac956}, \href
  {https://ui.adsabs.harvard.edu/abs/2018ApJ...861...49H} {861, 49}

\bibitem[\protect\citeauthoryear{{Helmboldt}, {Walterbos}  \&
  {Goto}}{{Helmboldt} et~al.}{2008}]{Helmboldt+2008}
{Helmboldt} J.~F.,  {Walterbos} R.~A.~M.,   {Goto} T.,  2008, \mn@doi [\mnras]
  {10.1111/j.1365-2966.2008.13229.x}, \href
  {https://ui.adsabs.harvard.edu/abs/2008MNRAS.387.1537H} {387, 1537}

\bibitem[\protect\citeauthoryear{{Hinz}, {Rieke}  \& {Caldwell}}{{Hinz}
  et~al.}{2003}]{Hinz+2003}
{Hinz} J.~L.,  {Rieke} G.~H.,   {Caldwell} N.,  2003, \mn@doi [\aj]
  {10.1086/379555}, \href
  {https://ui.adsabs.harvard.edu/abs/2003AJ....126.2622H} {126, 2622}

\bibitem[\protect\citeauthoryear{{Houghton}}{{Houghton}}{2015}]{Hou15}
{Houghton} R.~C.~W.,  2015, \mn@doi [Monthly Notices of the Royal Astronomical
  Society] {10.1093/mnras/stv1113}, \href
  {https://ui.adsabs.harvard.edu/abs/2015MNRAS.451.3427H} {451, 3427}

\bibitem[\protect\citeauthoryear{{Hubble}}{{Hubble}}{1936}]{Hubble+1936}
{Hubble} E.~P.,  1936, {Realm of the Nebulae}

\bibitem[\protect\citeauthoryear{{Huchra}, {Geller}  \& {Corwin}}{{Huchra}
  et~al.}{1995}]{Huchra+1995}
{Huchra} J.~P.,  {Geller} M.~J.,   {Corwin} Harold~G. J.,  1995, \mn@doi
  [\apjs] {10.1086/192191}, \href
  {https://ui.adsabs.harvard.edu/abs/1995ApJS...99..391H} {99, 391}

\bibitem[\protect\citeauthoryear{{Jones}, {Davies}  \& {Trewhella}}{{Jones}
  et~al.}{1996}]{JDT96}
{Jones} H.,  {Davies} J.~I.,   {Trewhella} M.,  1996, \mn@doi [\mnras]
  {10.1093/mnras/283.1.316}, \href
  {https://ui.adsabs.harvard.edu/abs/1996MNRAS.283..316J} {283, 316}

\bibitem[\protect\citeauthoryear{{Jones} et~al.,}{{Jones}
  et~al.}{2004}]{Jones+2004}
{Jones} D.~H.,  et~al., 2004, \mn@doi [\mnras]
  {10.1111/j.1365-2966.2004.08353.x}, \href
  {https://ui.adsabs.harvard.edu/abs/2004MNRAS.355..747J} {355, 747}

\bibitem[\protect\citeauthoryear{{Kennicutt} \& {Evans}}{{Kennicutt} \&
  {Evans}}{2012}]{KE12}
{Kennicutt} R.~C.,  {Evans} N.~J.,  2012, \mn@doi [\araa]
  {10.1146/annurev-astro-081811-125610}, \href
  {https://ui.adsabs.harvard.edu/abs/2012ARA&A..50..531K} {50, 531}

\bibitem[\protect\citeauthoryear{{Masters} et~al.,}{{Masters}
  et~al.}{2010}]{Masters+10}
{Masters} K.~L.,  et~al., 2010, \mn@doi [\mnras]
  {10.1111/j.1365-2966.2010.16335.x}, \href
  {https://ui.adsabs.harvard.edu/abs/2010MNRAS.404..792M} {404, 792}

\bibitem[\protect\citeauthoryear{{Mishra}, {Barway}  \& {Wadadekar}}{{Mishra}
  et~al.}{2017}]{Mishra+2017}
{Mishra} P.~K.,  {Barway} S.,   {Wadadekar} Y.,  2017, \mn@doi [\mnras]
  {10.1093/mnrasl/slx142}, \href
  {https://ui.adsabs.harvard.edu/abs/2017MNRAS.472L..89M} {472, L89}

\bibitem[\protect\citeauthoryear{{Moore}, {Katz}, {Lake}, {Dressler}  \&
  {Oemler}}{{Moore} et~al.}{1996}]{Moore+1996}
{Moore} B.,  {Katz} N.,  {Lake} G.,  {Dressler} A.,   {Oemler} A.,  1996,
  \mn@doi [\nat] {10.1038/379613a0}, \href
  {https://ui.adsabs.harvard.edu/abs/1996Natur.379..613M} {379, 613}

\bibitem[\protect\citeauthoryear{{Morganti} et~al.,}{{Morganti}
  et~al.}{2006}]{Mor06}
{Morganti} R.,  et~al., 2006, \mn@doi [\mnras]
  {10.1111/j.1365-2966.2006.10681.x}, \href
  {https://ui.adsabs.harvard.edu/abs/2006MNRAS.371..157M} {371, 157}

\bibitem[\protect\citeauthoryear{{Nair} \& {Abraham}}{{Nair} \&
  {Abraham}}{2010}]{Nair&Abraham2010}
{Nair} P.~B.,  {Abraham} R.~G.,  2010, \mn@doi [\apjs]
  {10.1088/0067-0049/186/2/427}, \href
  {https://ui.adsabs.harvard.edu/abs/2010ApJS..186..427N} {186, 427}

\bibitem[\protect\citeauthoryear{Otsu}{Otsu}{1979}]{Otsu+1979}
Otsu N.,  1979, IEEE transactions on systems, man, and cybernetics, 9, 62

\bibitem[\protect\citeauthoryear{{Poggianti}, {Smail}, {Dressler}, {Couch},
  {Barger}, {Butcher}, {Ellis}  \& {Oemler}}{{Poggianti}
  et~al.}{1999}]{Poggianti+1999}
{Poggianti} B.~M.,  {Smail} I.,  {Dressler} A.,  {Couch} W.~J.,  {Barger}
  A.~J.,  {Butcher} H.,  {Ellis} R.~S.,   {Oemler} Augustus J.,  1999, \mn@doi
  [\apj] {10.1086/307322}, \href
  {https://ui.adsabs.harvard.edu/abs/1999ApJ...518..576P} {518, 576}

\bibitem[\protect\citeauthoryear{{Postman} \& {Geller}}{{Postman} \&
  {Geller}}{1984}]{PG84}
{Postman} M.,  {Geller} M.~J.,  1984, \mn@doi [The Astrophysical Journal]
  {10.1086/162078}, \href
  {https://ui.adsabs.harvard.edu/abs/1984ApJ...281...95P} {281, 95}

\bibitem[\protect\citeauthoryear{{Querejeta} et~al.,}{{Querejeta}
  et~al.}{2015}]{Querejeta+2015}
{Querejeta} M.,  et~al., 2015, \mn@doi [\aap] {10.1051/0004-6361/201526354},
  \href {https://ui.adsabs.harvard.edu/abs/2015A&A...579L...2Q} {579, L2}

\bibitem[\protect\citeauthoryear{{Rines}, {Geller}, {Kurtz}  \&
  {Diaferio}}{{Rines} et~al.}{2005}]{Rines+2005}
{Rines} K.,  {Geller} M.~J.,  {Kurtz} M.~J.,   {Diaferio} A.,  2005, \mn@doi
  [\aj] {10.1086/433173}, \href
  {https://ui.adsabs.harvard.edu/abs/2005AJ....130.1482R} {130, 1482}

\bibitem[\protect\citeauthoryear{{Ronen}, {Aragon-Salamanca}  \&
  {Lahav}}{{Ronen} et~al.}{1999}]{Ronen+1999}
{Ronen} S.,  {Aragon-Salamanca} A.,   {Lahav} O.,  1999, \mn@doi [\mnras]
  {10.1046/j.1365-8711.1999.02222.x}, \href
  {https://ui.adsabs.harvard.edu/abs/1999MNRAS.303..284R} {303, 284}

\bibitem[\protect\citeauthoryear{{Salim} et~al.,}{{Salim}
  et~al.}{2016}]{Salim+2016}
{Salim} S.,  et~al., 2016, \mn@doi [\apjs] {10.3847/0067-0049/227/1/2}, \href
  {https://ui.adsabs.harvard.edu/abs/2016ApJS..227....2S} {227, 2}

\bibitem[\protect\citeauthoryear{{Salim}, {Boquien}  \& {Lee}}{{Salim}
  et~al.}{2018}]{Salim+2018}
{Salim} S.,  {Boquien} M.,   {Lee} J.~C.,  2018, \mn@doi [\apj]
  {10.3847/1538-4357/aabf3c}, \href
  {https://ui.adsabs.harvard.edu/abs/2018ApJ...859...11S} {859, 11}

\bibitem[\protect\citeauthoryear{{Sandage}}{{Sandage}}{2005}]{Sandage2005}
{Sandage} A.,  2005, \mn@doi [\araa] {10.1146/annurev.astro.43.112904.104839},
  \href {https://ui.adsabs.harvard.edu/abs/2005ARA&A..43..581S} {43, 581}

\bibitem[\protect\citeauthoryear{{Solanes}, {Salvador-Sol\'e}  \&
  {Sanrom\`a}}{{Solanes} et~al.}{1989}]{Solanes+1989}
{Solanes} J.~M.,  {Salvador-Sol\'e} E.,   {Sanrom\`a} M.,  1989, \mn@doi [\aj]
  {10.1086/115179}, \href {http://adsabs.harvard.edu/abs/1989AJ.....98..798S}
  {98, 798}

\bibitem[\protect\citeauthoryear{{Solanes}, {Giovanelli}  \&
  {Haynes}}{{Solanes} et~al.}{1996}]{Solanes+1996}
{Solanes} J.~M.,  {Giovanelli} R.,   {Haynes} M.~P.,  1996, \mn@doi [\apj]
  {10.1086/177089}, \href
  {https://ui.adsabs.harvard.edu/\#abs/1996ApJ...461..609S} {461, 609}

\bibitem[\protect\citeauthoryear{{Spitzer} \& {Baade}}{{Spitzer} \&
  {Baade}}{1951}]{Spitzer&Baade1951}
{Spitzer} Lyman J.,  {Baade} W.,  1951, \mn@doi [\apj] {10.1086/145406}, \href
  {https://ui.adsabs.harvard.edu/abs/1951ApJ...113..413S} {113, 413}

\bibitem[\protect\citeauthoryear{{Strauss} et~al.,}{{Strauss}
  et~al.}{2002}]{Strauss+2002}
{Strauss} M.~A.,  et~al., 2002, \mn@doi [\aj] {10.1086/342343}, \href
  {https://ui.adsabs.harvard.edu/abs/2002AJ....124.1810S} {124, 1810}

\bibitem[\protect\citeauthoryear{{Suzuki} et~al.,}{{Suzuki}
  et~al.}{2016}]{Suzuki+2016}
{Suzuki} T.~L.,  et~al., 2016, \mn@doi [\mnras] {10.1093/mnras/stw1655}, \href
  {https://ui.adsabs.harvard.edu/abs/2016MNRAS.462..181S} {462, 181}

\bibitem[\protect\citeauthoryear{{Tapia}, {Eliche-Moral}, {Aceves},
  {Rodr{\'\i}guez-P{\'e}rez}, {Borlaff}  \& {Querejeta}}{{Tapia}
  et~al.}{2017}]{Tapia+2017}
{Tapia} T.,  {Eliche-Moral} M.~C.,  {Aceves} H.,  {Rodr{\'\i}guez-P{\'e}rez}
  C.,  {Borlaff} A.,   {Querejeta} M.,  2017, \mn@doi [\aap]
  {10.1051/0004-6361/201628821}, \href
  {https://ui.adsabs.harvard.edu/abs/2017A&A...604A.105T} {604, A105}

\bibitem[\protect\citeauthoryear{{Tempel}, {Tuvikene}, {Kipper}  \&
  {Libeskind}}{{Tempel} et~al.}{2017}]{Tempel+2017}
{Tempel} E.,  {Tuvikene} T.,  {Kipper} R.,   {Libeskind} N.~I.,  2017, \mn@doi
  [\aap] {10.1051/0004-6361/201730499}, \href
  {https://ui.adsabs.harvard.edu/abs/2017A&A...602A.100T} {602, A100}

\bibitem[\protect\citeauthoryear{{Thomas} et~al.,}{{Thomas}
  et~al.}{2013}]{Thomas+13}
{Thomas} D.,  et~al., 2013, \mn@doi [\mnras] {10.1093/mnras/stt261}, \href
  {https://ui.adsabs.harvard.edu/abs/2013MNRAS.431.1383T} {431, 1383}

\bibitem[\protect\citeauthoryear{{Treu}, {Ellis}, {Kneib}, {Dressler}, {Smail},
  {Czoske}, {Oemler}  \& {Natarajan}}{{Treu} et~al.}{2003}]{Tre03}
{Treu} T.,  {Ellis} R.~S.,  {Kneib} J.-P.,  {Dressler} A.,  {Smail} I.,
  {Czoske} O.,  {Oemler} A.,   {Natarajan} P.,  2003, \mn@doi [The
  Astrophysical Journal] {10.1086/375314}, \href
  {https://ui.adsabs.harvard.edu/abs/2003ApJ...591...53T} {591, 53}

\bibitem[\protect\citeauthoryear{{Trouille} \& {Barger}}{{Trouille} \&
  {Barger}}{2010}]{Trouille+Barger2010}
{Trouille} L.,  {Barger} A.~J.,  2010, \mn@doi [\apj]
  {10.1088/0004-637X/722/1/212}, \href
  {https://ui.adsabs.harvard.edu/abs/2010ApJ...722..212T} {722, 212}

\bibitem[\protect\citeauthoryear{{Welch} \& {Sage}}{{Welch} \&
  {Sage}}{2003}]{WS03}
{Welch} G.~A.,  {Sage} L.~J.,  2003, \mn@doi [\apj] {10.1086/345537}, \href
  {https://ui.adsabs.harvard.edu/abs/2003ApJ...584..260W} {584, 260}

\bibitem[\protect\citeauthoryear{{Willett} et~al.,}{{Willett}
  et~al.}{2013}]{Willett+2013}
{Willett} K.~W.,  et~al., 2013, \mn@doi [\mnras] {10.1093/mnras/stt1458}, \href
  {https://ui.adsabs.harvard.edu/abs/2013MNRAS.435.2835W} {435, 2835}

\bibitem[\protect\citeauthoryear{{Williams}, {Bureau}  \&
  {Cappellari}}{{Williams} et~al.}{2010}]{Williams+2010}
{Williams} M.~J.,  {Bureau} M.,   {Cappellari} M.,  2010, \mn@doi [\mnras]
  {10.1111/j.1365-2966.2010.17406.x}, \href
  {https://ui.adsabs.harvard.edu/abs/2010MNRAS.409.1330W} {409, 1330}

\bibitem[\protect\citeauthoryear{{Xiao}, {Gu}, {Chen}  \& {Zhou}}{{Xiao}
  et~al.}{2016}]{Xiao+2016}
{Xiao} M.-Y.,  {Gu} Q.-S.,  {Chen} Y.-M.,   {Zhou} L.,  2016, \mn@doi [\apj]
  {10.3847/0004-637X/831/1/63}, \href
  {https://ui.adsabs.harvard.edu/abs/2016ApJ...831...63X} {831, 63}

\bibitem[\protect\citeauthoryear{{Yip} et~al.,}{{Yip} et~al.}{2004}]{Yip}
{Yip} C.~W.,  et~al., 2004, \mn@doi [\aj] {10.1086/422429}, \href
  {https://ui.adsabs.harvard.edu/abs/2004AJ....128..585Y} {128, 585}

\bibitem[\protect\citeauthoryear{{York} et~al.,}{{York}
  et~al.}{2000}]{York+2000}
{York} D.~G.,  et~al., 2000, \mn@doi [\aj] {10.1086/301513}, \href
  {https://ui.adsabs.harvard.edu/abs/2000AJ....120.1579Y} {120, 1579}

\bibitem[\protect\citeauthoryear{{de Vaucouleurs}}{{de
  Vaucouleurs}}{1977}]{Vaucouleurs+1977}
{de Vaucouleurs} G.,  1977, in {Tinsley} B.~M.,  {Larson} Richard B.~Gehret
  D.~C.,  eds, Evolution of Galaxies and Stellar Populations. p.~43

\bibitem[\protect\citeauthoryear{{den Heijer} et~al.,}{{den Heijer}
  et~al.}{2015}]{denHeijer+2015}
{den Heijer} M.,  et~al., 2015, \mn@doi [\aap] {10.1051/0004-6361/201526879},
  \href {https://ui.adsabs.harvard.edu/abs/2015A&A...581A..98D} {581, A98}

\bibitem[\protect\citeauthoryear{{van den Bergh}}{{van den
  Bergh}}{2009}]{vdB2009}
{van den Bergh} S.,  2009, \mn@doi [\apj] {10.1088/0004-637X/702/2/1502}, \href
  {https://ui.adsabs.harvard.edu/abs/2009ApJ...702.1502V} {702, 1502}

\makeatother
\end{thebibliography}


\onecolumn
\appendix
\newpage

\section{Testing the dependence of the results on the observational set-up}
\label{app:robustness}

In this appendix, we investigate whether there are factors alien to the population of lenticulars that could be biasing their spectral properties and, hence, their classification and other outcomes of this work. 

\subsection{Effects of aperture: central-to-total light contribution}
\label{sapp:aperture}

Our analysis of nearby, bulge-enhanced S0 galaxies is based on SDSS optical galaxy spectra taken using single fibres. The relatively small diameter of these fibres (3 arcsec) and moderate depth of the selected dataset imply that the flux distributions they collect do not come from the galaxies as a whole, but from their central regions. For the nearest lenticulars, this could cause the light entering the spectrograph to essentially come from their bulge component. Were it to be the case, and assuming bulges are generally a passive element no co-evolving with discs \citetext{but see \citealt{mckelvie+2018}}, one might expect to see a correlation between the apparent size of the galaxies and their spectra. Therefore, it is not unreasonable to assume that our spectral classification into passive and active lenticulars could be affected by artefacts associated with the fixed fibre aperture, since the sampled physical scales vary significantly (from $\sim 0.405$ to $1.85$ kpc/arcsec for redshifts between $0.02$ and $0.1$, respectively).

\begin{center}
	\includegraphics[width=0.575\columnwidth]{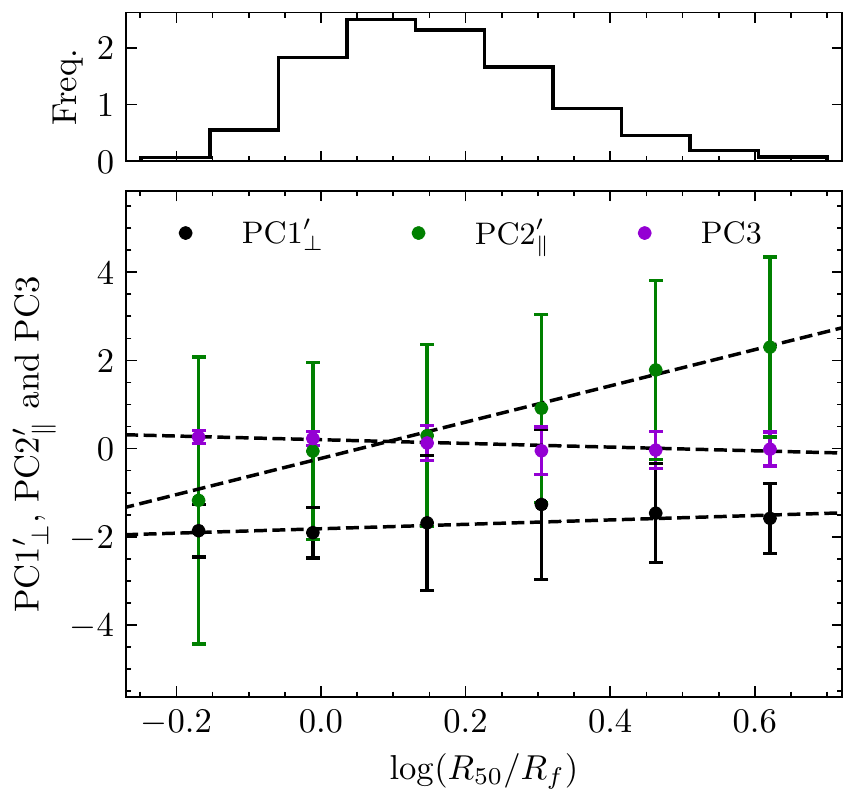}
    \captionof{figure}{\small Top: Frequency distribution of the observed apparent sizes of S0 galaxies with respect to the radius of the SDSS fibres. Bottom: Median values and interquartile ranges of the three main PC scores for the MLSS0 in bins of $\log(R_{50}/R_{\mathrm f})$ rotated about the PC3 axis so that the PS for the MLSS0 becomes parallel to the $Y$ axis (see the text for details). The dashed straight lines correspond to the linear laws fitted to the data points in the new base (black circles for the new abscissae, PC1$^\prime_\bot$, green for the new ordinates, PC2$^\prime_\parallel$, and purple for the unaffected PC3 values). They can be used to correct, from a statistical point of view, the S0's spectra for aperture effects. The nearly horizontal line associated with PC1$^\prime_\bot$ demonstrates that this correction would move the PC scores along a direction essentially parallel to the PS.}
    \label{fig:test2}   
\end{center}

To estimate the potential bias this may induce in the spectral classification, we have devised a strategy consisting first in quantifying the effects of the fixed fibre aperture on the spectra, or equivalently, on the projections of such spectra on to the principal components. This has been done by calculating the dependence of PC1, PC2, and PC3, for the entire S0 population, on $\log(R_{50}/R_{\mathrm f})$, the logarithm of the ratio between the Petrosian-based estimate of the projected half-light radius in the $r$-band provided by the NSA photometry and the radius of the SDSS fibres. The inferred variations turn out to be very well approximated by linear laws. These simple relationships facilitate the derivation of a straightforward statistical correction to the data in the subspace of the main principal components to compensate for the aperture effects\footnote{Since any spectrum can be represented by an arbitrarily large linear combination of the ES weighted by the corresponding PC, this statistical correction could also be directly applied to the original spectra.} (see below), while revealing that such compensation entails shifts of different magnitude in the PC values along a common direction.

As shown in the bottom panel of Fig.~\ref{fig:test2}, the direction of those shifts is nearly parallel to the PS sequence. This plot depicts the median values of the first three PC scores per bin of aperture size after applying a clockwise rotation of $\sim 40^\circ$ around the PC3 axis. This is the angle subtended by the observed PS ridge of the MLSS0 with the original PC2 axis (see the right panel of Fig.~\ref{fig:split}), which has been inferred from an orthogonal regression fit to the data. In this new basis the PS viewed along the direction of the PC3 axis adopts an upright position with the AC extending horizontally, so the names assigned to the new axes: PC1$^\prime_\bot$, PC2$^\prime_\parallel$, and PC3, respectively\footnote{We have permitted ourselves to abuse the language since arbitrary rotations do not preserve the PC condition, hence the prime superscript.}. In the rotated frame of reference, the statistical correction of the aperture effects can be expressed in vector form as
\begin{equation}
    (\text{PC1}^\prime_\bot,\text{PC2}^\prime_\parallel,\text{PC3})^{\text{cor}} = (\text{PC1}^\prime_\bot,\text{PC2}^\prime_\parallel,\text{PC3})^{\text{obs}}+[\log(R_{50}/R_{\mathrm f})^* - \log(R_{50}/R_{\mathrm f})]\cdot(m_1,m_2,m_3)\;,
    \label{eq:corr_aperture}
\end{equation}
where $\log(R_{50}/R_{\mathrm f})^*$ represents any arbitrary reference value that one wishes to adopt for this quantity -- it could be, for instance, the value of the most populated bin --, and $m_1 = 0.40$, $m_2 = 3.7$, and $m_3 = -0.30$ are the slopes of the fitted straight lines. As shown by the figure, the correction of the aperture effects would move the PC scores in a direction essentially parallel to the vertical axis that now is also the PS direction: it forms an angle of less than $6^\circ$! More specifically, we have calculated that the application of this correction would imply the reclassification of a tiny fraction (always less than one per cent) of the S0 galaxies between adjacent spectral classes, whereas there would not be a single case of change of status between PS and AC. Therefore, despite the evident effects of the fixed aperture of the fibres on the S0's spectra, it is clear that their impact on the PS/TR/AC classification is practically insignificant, so there is no need to take corrective actions in this regard.

The histogram in the upper panel of Fig.~\ref{fig:test2} provides another useful result related to this issue. According to the distribution of $B/D$ flux ratios shown by S0 \citetext{see, for example, the review by \citealt{Graham2014}}, the  bulge-to-total light ratios for this population should range from $\sim 0.1$ to $\sim 0.3$, with the peak at $0.2$. An eyeball estimation of the radius encompassing these light fractions that accounts for the typical differences in the bulge and disc light profiles indicates that one should expect the fibres to capture light essentially coming from the bulge component, even in highly inclined S0, for those galaxies on which $\log(R_{50}/R_{\mathrm f})\gtrsim 0.4$--$0.5$ (this is something that we have empirically confirmed). As illustrated by the histogram, there are very few objects verifying this condition. To give some figures by way of example: only a 3 percent of S0 have $\log(R_{50}/R_{\mathrm f}) > 0.5$. This implies that nearly all the single-fibre spectra evaluated should contain some light coming from the disc component.

\subsection{Effects of inclination: internal dust reddening}
\label{sapp:inclination}

The apparent inclination of the galaxies is another factor that might lead to their spectral misclassification. For a given galaxy at a fixed distance it determines the relative contributions to the spectrum of the different structural components of the galaxy, which in turn are modulated by the degree of its internal extinction. In fact, the detection in the present work of a substantial fraction of S0 with obvious signs of ongoing star formation, as well as the well-known fact that even lenticulars that have used up or lost most of their interstellar matter may retain significant dust in their discs, suggest that one may expect internal absorption in these objects to be not totally negligible, especially for highly-inclined star-forming AC S0 (recall that we have already argued, based on Figs.~\ref{fig:Avvsba} and \ref{fig:inclination_hist}, that this sub-population should be moderately opaque).

\begin{center}
	\includegraphics[width=\textwidth]{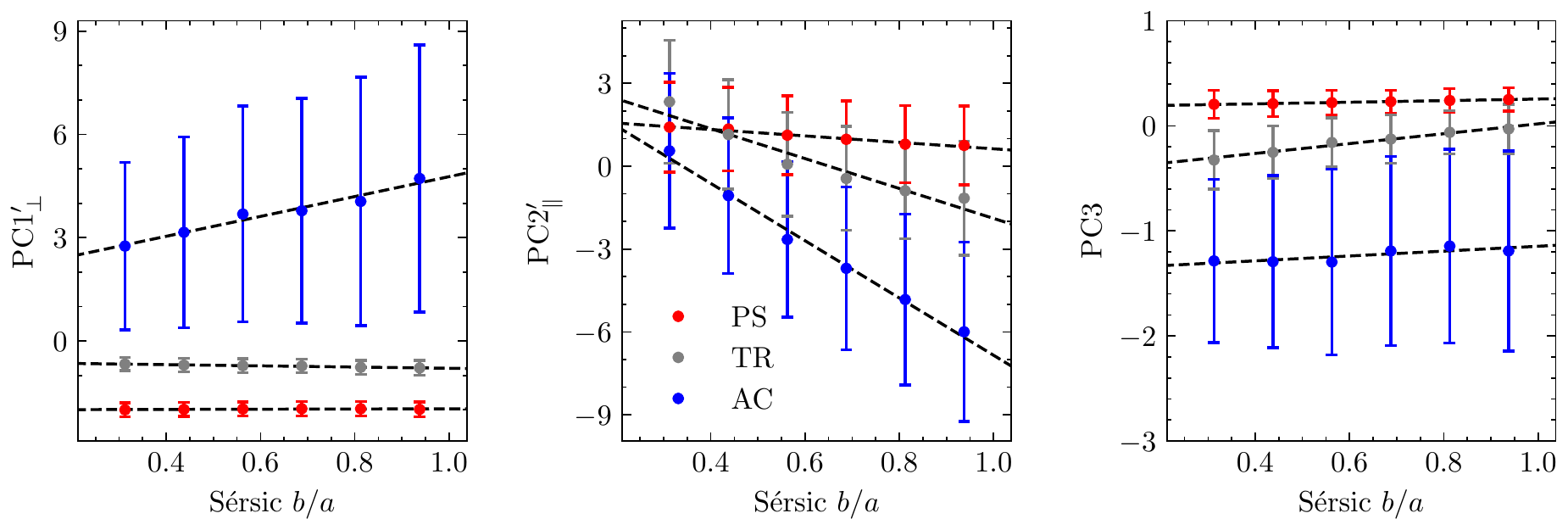}
    \captionof{figure}{\small Median values and interquartile ranges of the three main PC scores for the PS (red circles), TR (grey circles), and AC (blue circles) galaxies of the MLSS0 in bins of the observed axial ratio $b/a$, with the latter inferred from 2D single-component S\'ersic fits in $r$-band. We show, using separate panels, the values of the main three scores in the same rotated frame of reference used in Fig.~\ref{fig:test2}. The dashed straight lines correspond to the linear laws fitted to the data points. They can be used to correct, from a statistical point of view, the S0's spectra for inclination effects. The plot illustrates an increasing sensibility of the spectral classes to inclination: being essentially null for the PS, mild for the TR, and somewhat more significant for the AC. As in other figures in which the ratio $b/a$ is also used, the intrinsic axial ratio of S0 galaxies has been taken into account when introducing a lower limit on the observed inclination values.}
    \label{fig:test1}   
\end{center}

This possibility is confirmed by Fig.~\ref{fig:test1}, where we show the results of applying the same procedure described in Appendix~\ref{sapp:aperture} that transforms the first three PC to a new base in which the PS becomes parallel to the $Y$ axis, to the PS, TR, and AC spectral classes separately. This plot reveals that the values of the two main PC of the AC galaxies and of the second one of TR objects are the most sensitive to changes on inclination. It may also be observed that the strength of the bias tends to decrease with the order of the PC. At the other extreme, the impact of $b/a$ on the spectra of PS members is negligible, which is consistent with our appraisal that the galaxies of this latter class should contain very small amounts of gas and dust. 

As in the case of the aperture effects, the inclination dependencies of the transformed PC scores shown in the three panels of Fig.~\ref{fig:test1} are very well approximated by linear laws that can be used to derive a statistical correction to the S0 spectra that compensates for this bias. But again, we find that the correction of the inclination would move the original PC values in directions away from the uncorrected PS ridge that form small angles with it: $< 2^\circ$, $\sim 2^\circ$, and $\sim 15^\circ$, respectively, for the PS, TR and AC galaxies. This means that the only correction that could reasonably be expected to have some effect on the spectral classification of S0 galaxies would be that corresponding to the latter class. By taking the values of the face-on (i.e. $b/a = 1$) scores as the baseline, this correction in the rotated frame of reference is  
\begin{equation}
    (\text{PC1}^\prime_\bot,\text{PC2}^\prime_\parallel,\text{PC3})_{\text{AC}}^{\text{cor}} = (\text{PC1}^\prime_\bot,\text{PC2}^\prime_\parallel,\text{PC3})_{\text{AC}}^{\text{obs}}+(1 - b/a)\cdot(m_1,m_2,m_3)\;,
    \label{eq:vec_redd}
\end{equation}
where $m_1 = 2.9$, $m_2 = -10.4$, and $m_3 = 0.20$ are the slopes of the fitted straight lines to the AC data\footnote{We are ignoring here the fact that $b/a$ is not fully equivalent to $\cos i$ at high inclinations $i$ due to the finite thickness of the lenticular galaxies.}. But much as with the aperture effects, the inclination biases translate in practice into mis-classification frequencies between adjacent spectral classes that are very small (of a few per cent at most) and, most importantly, into a null probability that PS galaxies will become AC and vice versa. It can therefore be concluded that inclination effects, and their correction, do not have a significant impact on the main conclusions of this work. Note also that corrections of the sort outlined by Equations~(\ref{eq:corr_aperture}) and (\ref{eq:vec_redd}) could be iterated if necessary until convergence.

Finally, it must be taken into account that the corrections reported above are specific to our sample. They must be recalculated for datasets of S0 spectra gathered with different selection rules, especially if they involve the use of other aperture sizes. In any case, the results of the present investigation suggest that, unless the differences in the observational constraints are dramatic, aperture and inclination effects present in single-fibre optical spectroscopic surveys of the local population of S0 galaxies appear to have a much more limited impact on their spectral classification than might be expected. Even so, the most efficient way to overcome these limitations is through IFS surveys, such as for instance the aforementioned MaNGA or the Sydney-AAO Multi-object Integral field spectrograph (SAMI) Galaxy Survey \citep{All15}, both currently under way. Censuses like these provide full two-dimensional spectral coverage of the galaxies and, once completed, will result in samples representative of the nearby universe that can easily be corrected to become volume-limited datasets.

\newpage
\section{Examples of spectra and images of members of the two spectral classes of S0}
\label{app:spectra}

\begin{center}
    \par\bigskip
	\includegraphics[width=0.48\columnwidth]{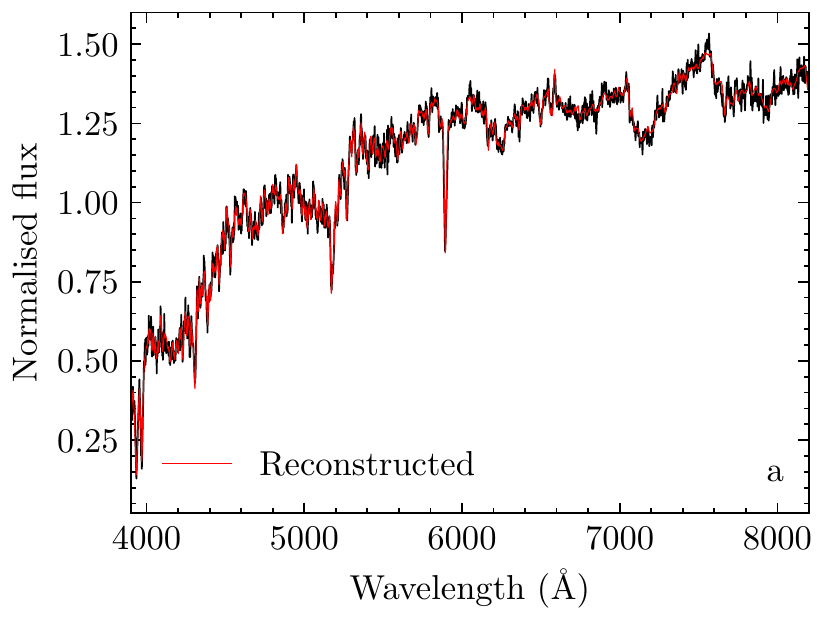}
	\includegraphics[width=0.48\columnwidth]{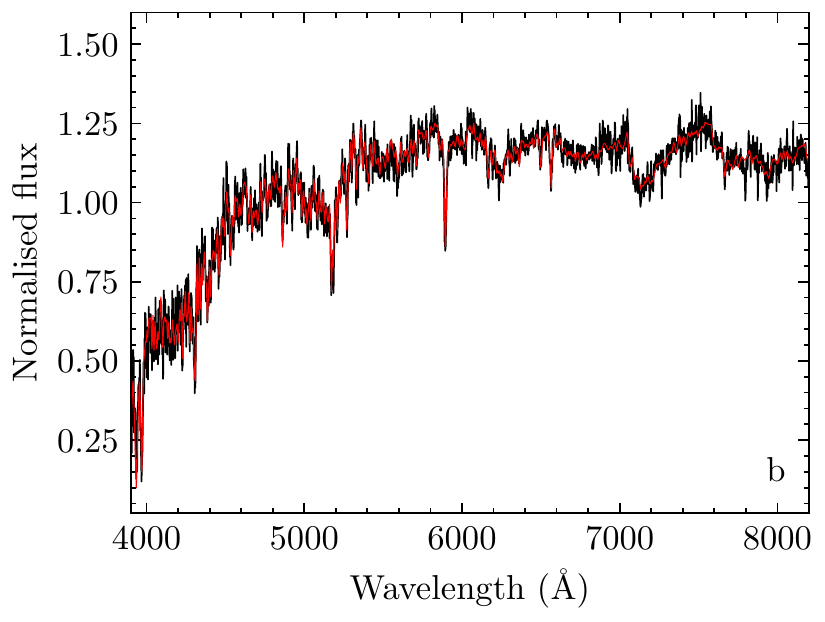}\par\bigskip
    \includegraphics[width=0.48\columnwidth]{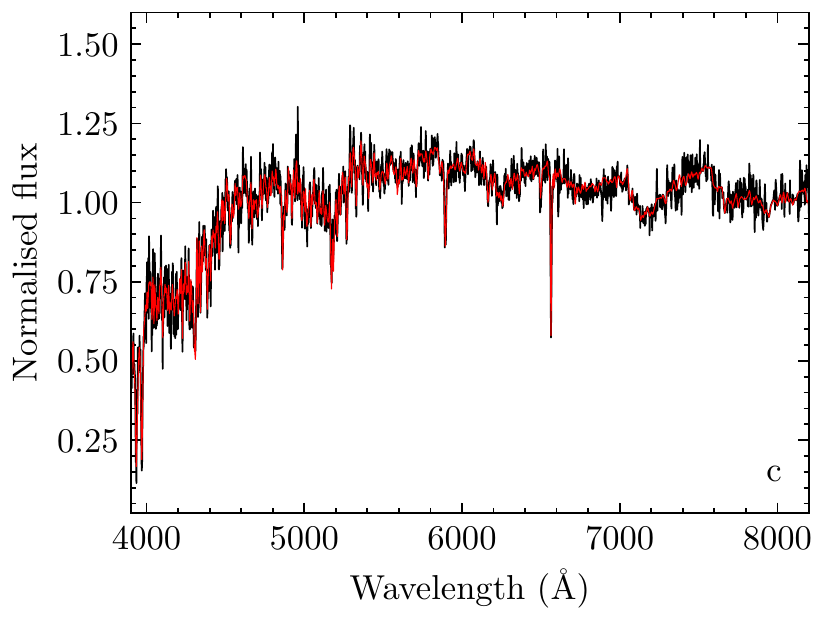}
	\includegraphics[width=0.48\columnwidth]{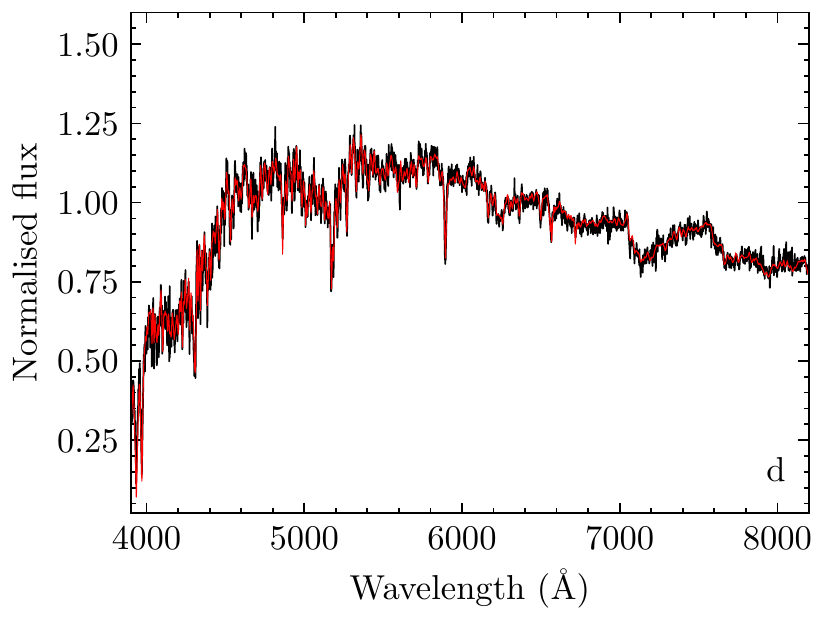}\par\bigskip
	\includegraphics[width=0.48\columnwidth]{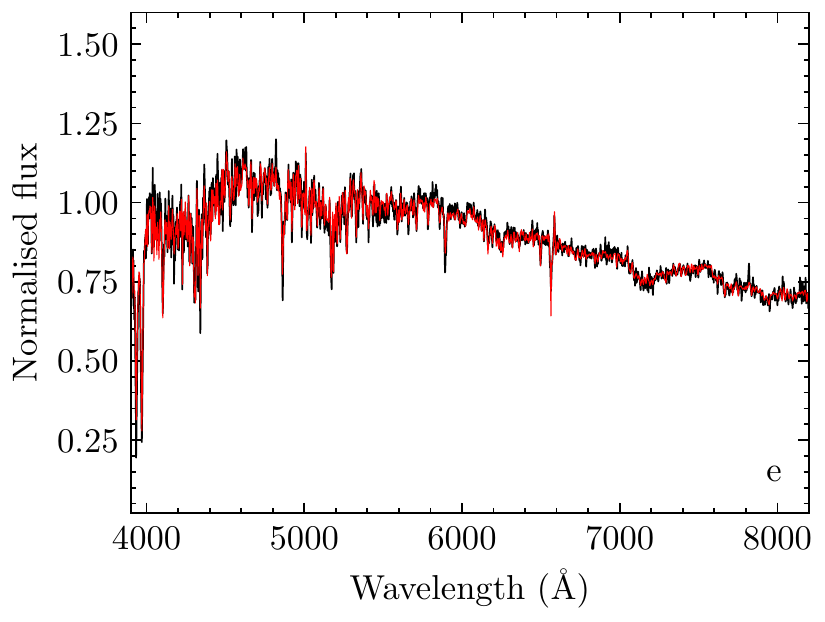} 
	\includegraphics[width=0.48\columnwidth]{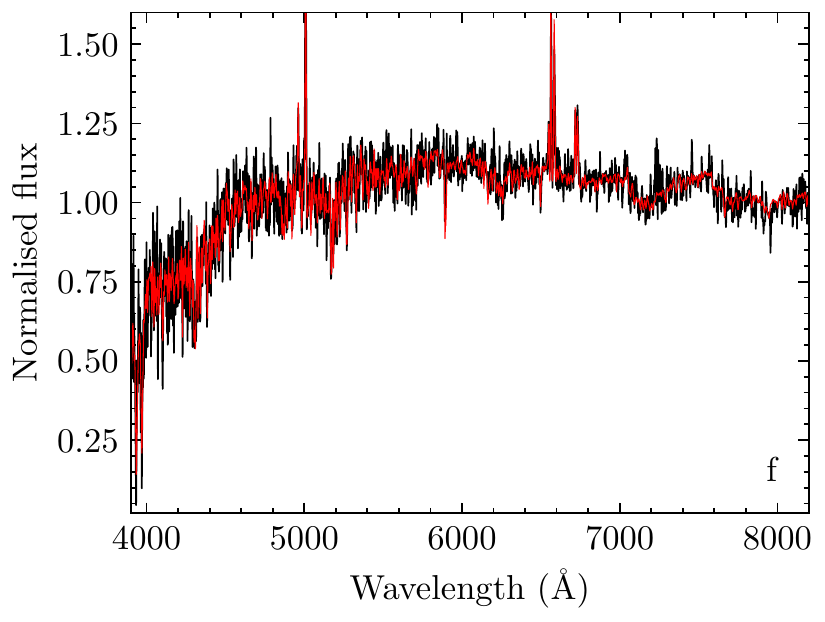}\par\smallskip
	\captionof{figure}{Random examples of individual S0 spectra at different positions of the PC1--PC2 subspace, marked in the last panel of this figure. Black curves are actual rest-frame shifted, re-binned, and normalised spectra of our main sample with gaps filled as explained in Section~\ref{S:pca}. The red curves on top show the reconstructed spectra obtained by combining the mean of the S0 spectra with the first ten eigenspectra, which explain $97$ per cent of the variance of the training sample (see Fig.~\ref{fig:samp_var}). Panels (a)--(e) are for galaxies included in the class dubbed Passive Sequence, whereas panels (f)--(k) are for members of the Active Cloud. The colour images of the galaxies producing these spectra are shown in Fig.~\ref{fig:img_samp}.}  
\end{center}	    

\setcounter{figure}{0}
\begin{figure*}
	\includegraphics[width=0.48\columnwidth]{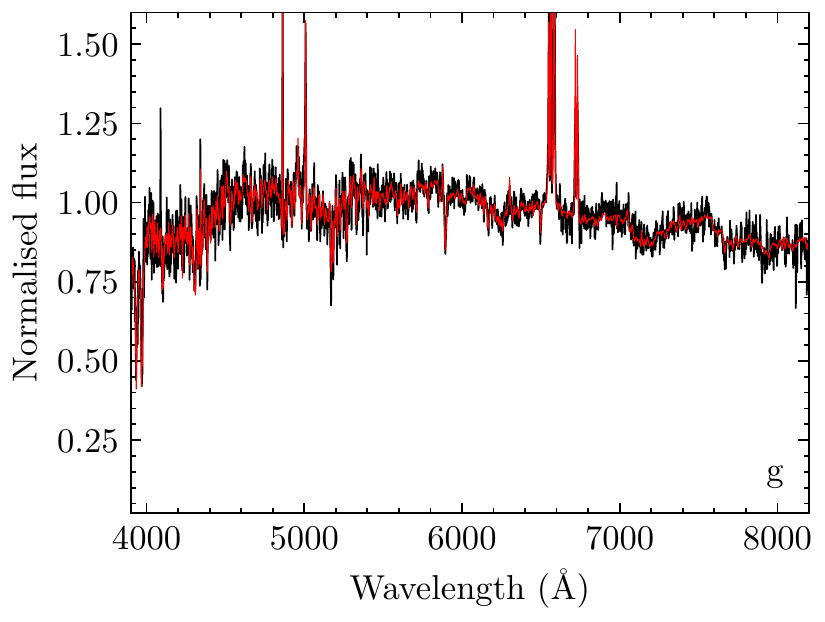} 
	\includegraphics[width=0.48\columnwidth]{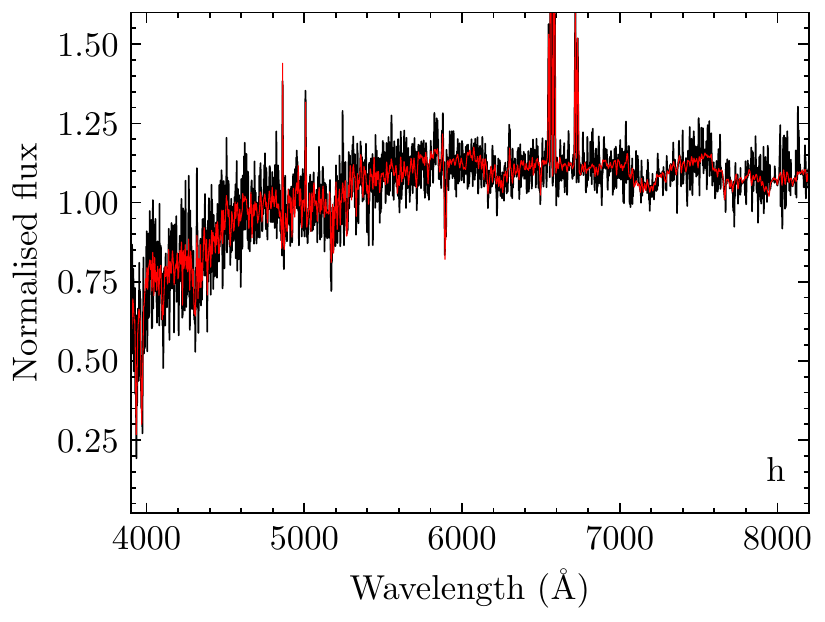}\par\bigskip
	\includegraphics[width=0.48\columnwidth]{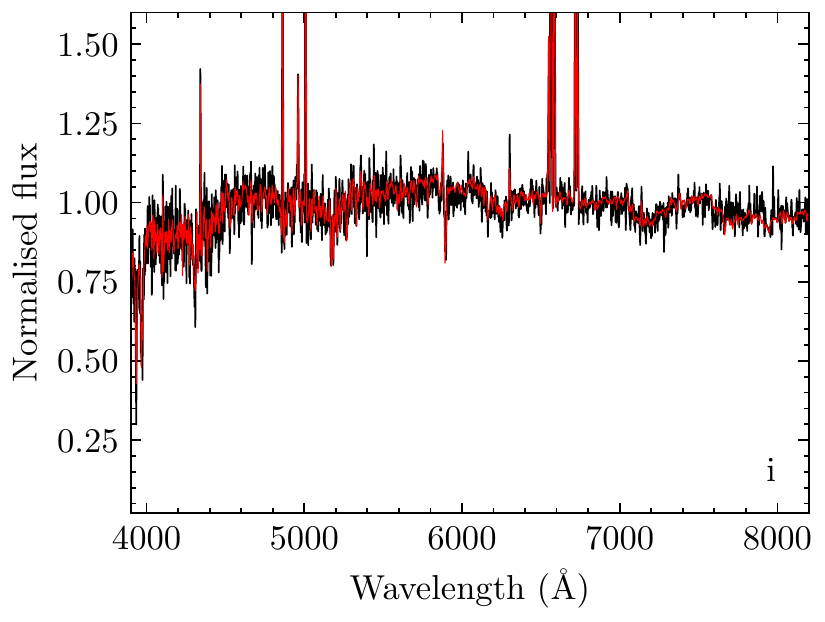}
	\includegraphics[width=0.48\columnwidth]{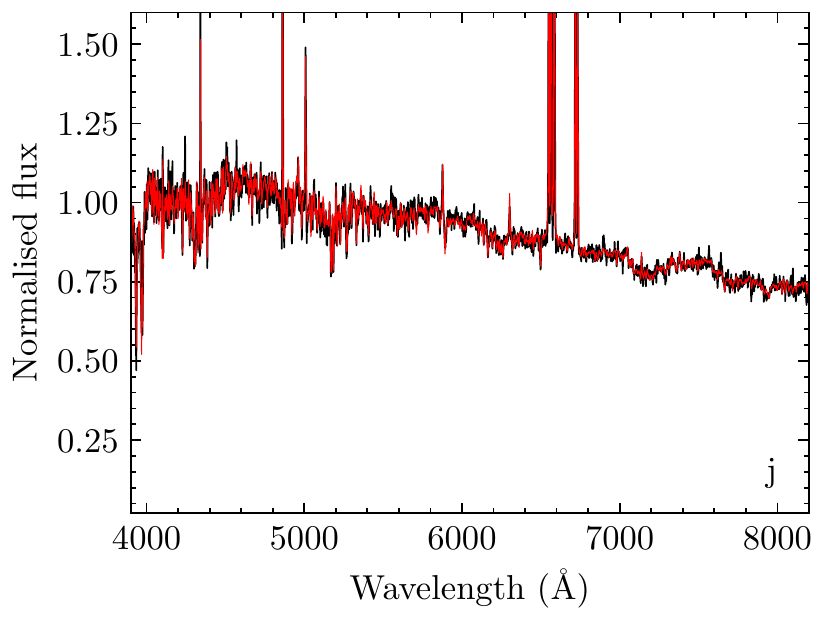}\par\bigskip
	\includegraphics[width=0.48\columnwidth]{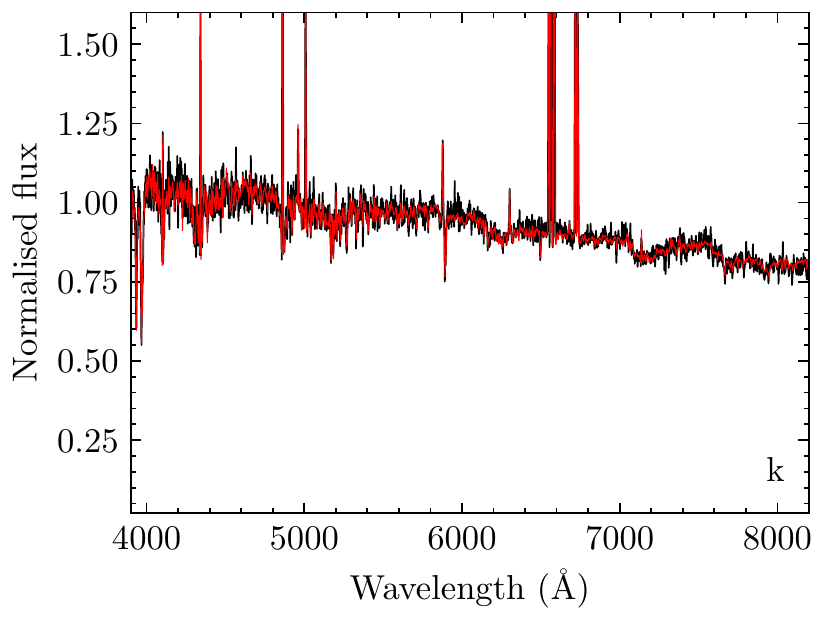}
	\includegraphics[width=0.48\columnwidth]{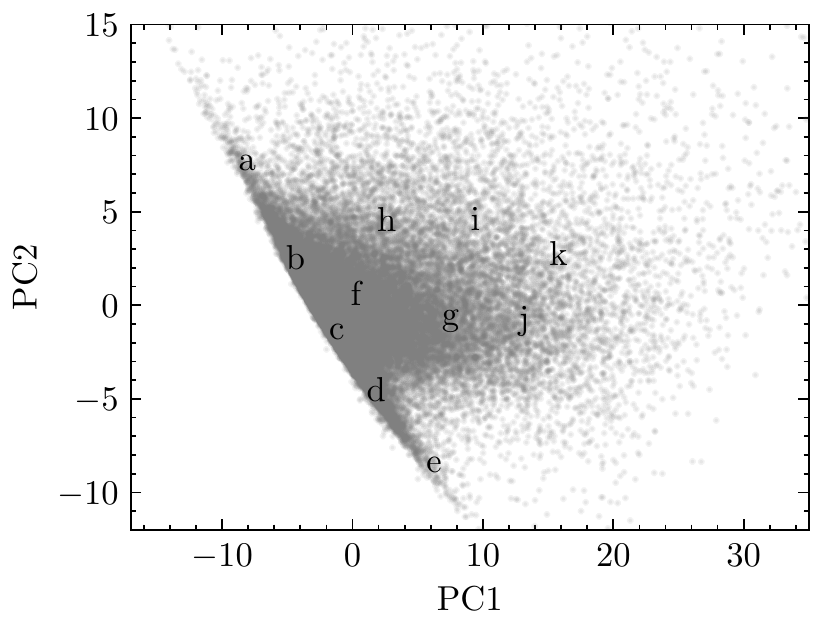} 
    \caption{Cont.}
\label{fig:spec_samp2}    
\end{figure*}
\vspace*{15\baselineskip}
\newpage
\begin{center}
	\includegraphics[width=\textwidth]{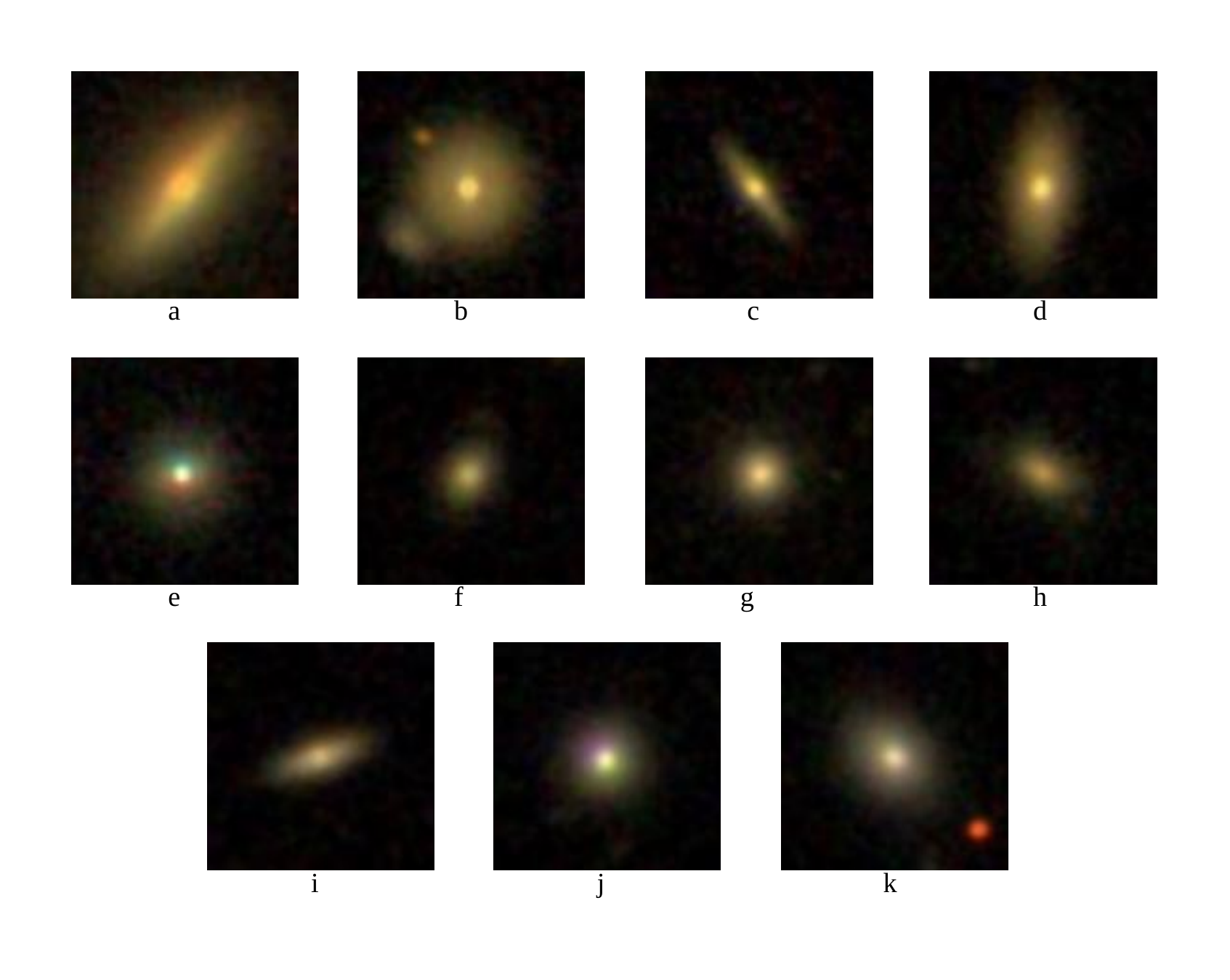}
	\vspace*{-2\baselineskip}
    \captionof{figure}{SDSS colour images of the S0 galaxies producing the spectra  shown in Fig.~\ref{fig:spec_samp2}.}
    \label{fig:img_samp}  
\end{center} 

\newpage
\section{PDF of some properties of the two spectral classes of S0}
\label{app:distributions}

\begin{center}
	\includegraphics[width=0.86\textwidth]{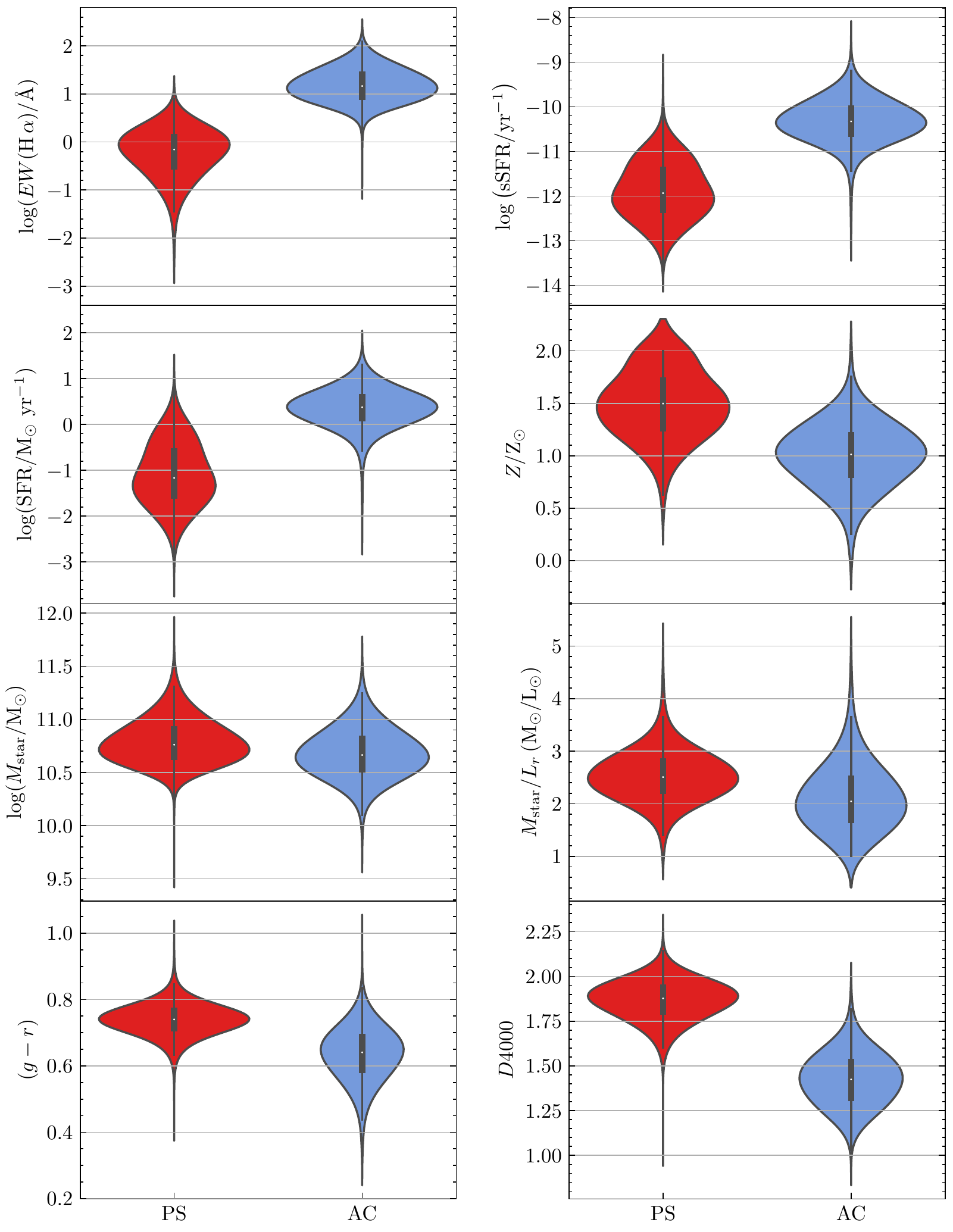}
    \captionof{figure}{Violin plots showing for the members of the Passive Sequence (red) and the Active Cloud (blue) in the VLSS0 their PDF for the eight spectrophotometric properties that are found to correlate most strongly with the first three PC. From left to right and top to bottom: $\text{H}\,\alpha$ equivalent width, specific star formation rate per unit stellar mass and per galaxy, metallicity and mass of the stellar population, stellar mass-to-$r$-band-light ratio, $(g-r)$ colour, and $D4000$ break. The central dots and boxes indicate, respectively, the medians and interquartile ranges of the distributions. In all cases the two-sample KS test returns p-values $\ll 0.05$ indicating a very low likelihood that the two main S0 spectral classes come from the same parent population.}   
\label{fig:violins}    
\end{center}


\bsp	
\label{lastpage}
\end{document}